\documentclass[aps,longbibliography,pre, reprint,groupaddress,bibnotes]{revtex4-1}

\usepackage{dcolumn}
\usepackage{bm, amssymb, color}
\usepackage{hyperref}
\usepackage{natbib}
\usepackage{booktabs,tabulary}
\usepackage{multirow}
\usepackage{amsmath,soul}
\usepackage[export]{adjustbox}
\usepackage{float}
\usepackage[caption=false]{subfig}
\usepackage{array}
\usepackage{cases}
\usepackage{makecell}
\usepackage{physics}
\usepackage{yhmath}

\usepackage{enumitem}

\newcommand{\spD}[1]{\fn{\tilde{\chi}_{_V}}{#1}}

\newcommand{\tens}[1]{\boldsymbol{#1}}
\newcommand{\ten}[2]{#1_{#2}}

\newcommand{\Hankel}[2]{\fn{\mathcal{H}_{#1}^{(1)}}{#2}}
\newcommand{\uvect}[1]{\hat{\vect{#1}}}

\setstcolor{red}

\setul{0}{0.3ex}

\newcommand{\R}{\mathbb{R}}

\newcommand{\fn}[2]{\mathinner{#1\mathopen{\left(#2\right)}}}
\newcommand{\vect}[1]{{\bf #1}}
\newcommand{\E}[1]{\left\langle#1\right\rangle}

\usepackage[colorinlistoftodos,shadow,textwidth=18mm]{todonotes}

\begin{document}

\title{Nonlocal Effective  Electromagnetic Wave Characteristics of Composite Media: Beyond the Quasistatic Regime}

\author{Salvatore Torquato}%
\email{torquato@princeton.edu}
\homepage{http://torquato.princeton.edu}
\affiliation{Department of Physics, Princeton University, Princeton, New Jersey 08544, USA}
\affiliation{Department of Chemistry, Princeton University, Princeton, New Jersey 08544, USA}
\affiliation{Princeton Institute for the Science and Technology of Materials, Princeton University, Princeton, New Jersey 08544, USA}
\affiliation{Program in Applied and Computational Mathematics, Princeton University, Princeton, New Jersey 08544, USA}
\author{Jaeuk Kim}
\affiliation{Department of Physics, Princeton University, Princeton, New Jersey 08544, USA}

\begin{abstract}
We derive exact nonlocal homogenized constitutive relations for the effective electromagnetic wave properties of disordered two-phase composites and metamaterials from first principles. 
This exact formalism enables us to extend the long-wavelength limitations of conventional homogenization estimates of the effective dynamic dielectric constant tensor $\tens{\varepsilon}_e(\vect{k}_I,\omega)$  for arbitrary microstructures so that it can capture spatial dispersion well beyond the quasistatic regime (where $\omega$ and $\vect{k}_I$ are the frequency and wave vector of the incident radiation).
We accomplish this task by deriving  nonlocal strong-contrast expansions that exactly account for complete microstructural information (infinite set of $n$-point correlation functions) and hence multiple scattering  
to all orders for the range of wave numbers for which our extended homogenization theory applies, i.e.,  $0 \le  |\vect{k}_I| \ell \lesssim 1$
(where $\ell$ is a characteristic heterogeneity length scale).
Because of the fast-convergence properties of such expansions, their lower-order truncations yield accurate closed-form approximate formulas for  $\tens{\varepsilon}_e(\vect{k}_I,\omega)$ that apply for a wide class of microstructures. 
These nonlocal formulas are resummed representations of the strong-contrast expansions that still accurately capture multiple scattering to all orders via the microstructural information embodied in the spectral density, which is easy to compute for any composite.
The accuracy of these microstructure-dependent approximations is validated by comparison to full-waveform simulation computations for both 2D and 3D ordered and disordered models of composite media. Thus, our closed-form formulas enable one to predict accurately and efficiently the effective wave characteristics well beyond the quasistatic regime for a wide class of composite microstructures without having to perform full-blown simulations. 
We find that disordered hyperuniform media are generally less lossy than their nonhyperuniform counterparts.
We also show that certain disordered hyperuniform particulate composites exhibit novel wave characteristics, including the capacity to act as low-pass filters that transmit waves
``isotropically” up to a selected wave number and refractive indices that abruptly change over a narrow range of wave numbers.
Our results demonstrate that one  can design the effective wave characteristics of a disordered composite by engineering 
the microstructure to possess tailored spatial correlations at prescribed length scales.
Thus, our findings can accelerate the discovery of novel electromagnetic composites.
\end{abstract}

\date{\today}

\maketitle

\section{Introduction}

The theoretical problem of estimating the effective properties of multiphase composite media is an outstanding one and dates back to work by some of the luminaries of science, including Maxwell \cite{Ma73}, Lord Rayleigh \cite{Ra92}, and Einstein \cite{Ei06}.
The preponderance of previous  theoretical studies have focused on the determination of static effective properties (e.g., dielectric constant, elastic moduli and fluid permeability) using a variety of methods, including approximation schemes \cite{Ma73,Br35,Br47,Bu65}, bounding techniques \cite{Pr61,Ha62c,Be65, Ko88b,To02a,Mi02} and exact series-expansion procedures \cite{Br55,Fe82a, Se89,To97b}.
The latter set of investigations teaches us that an exact determination of an effective property, given the phase properties of the composite, generally requires an infinite set of correlation functions that characterizes the composite microstructure.

\begin{figure*}[t]
\centering
\includegraphics[ height=4.5cm]{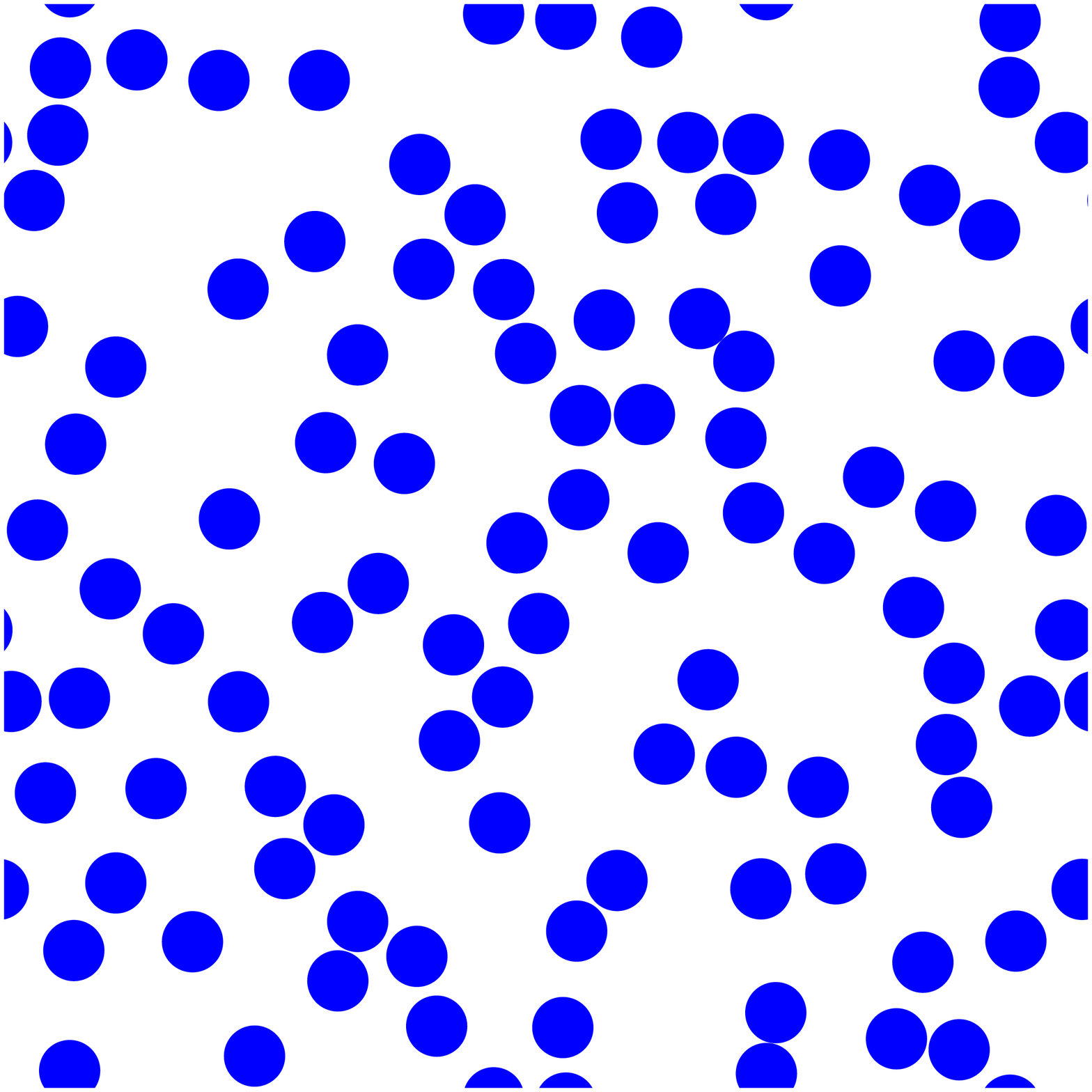}
\hspace{0.2in}
\includegraphics[height=4.5cm]{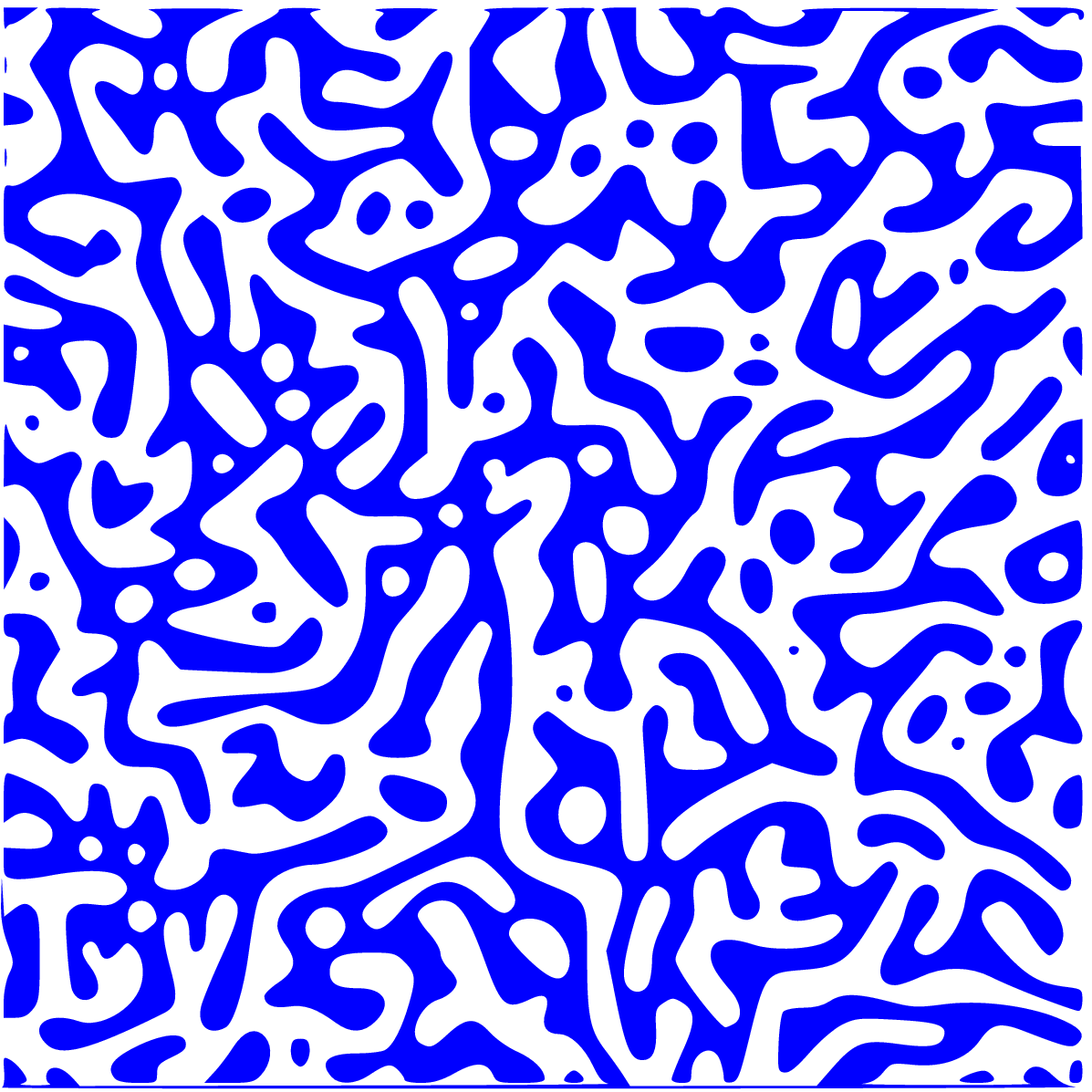}
\hspace{0.2in}
\includegraphics[height=4.5cm]{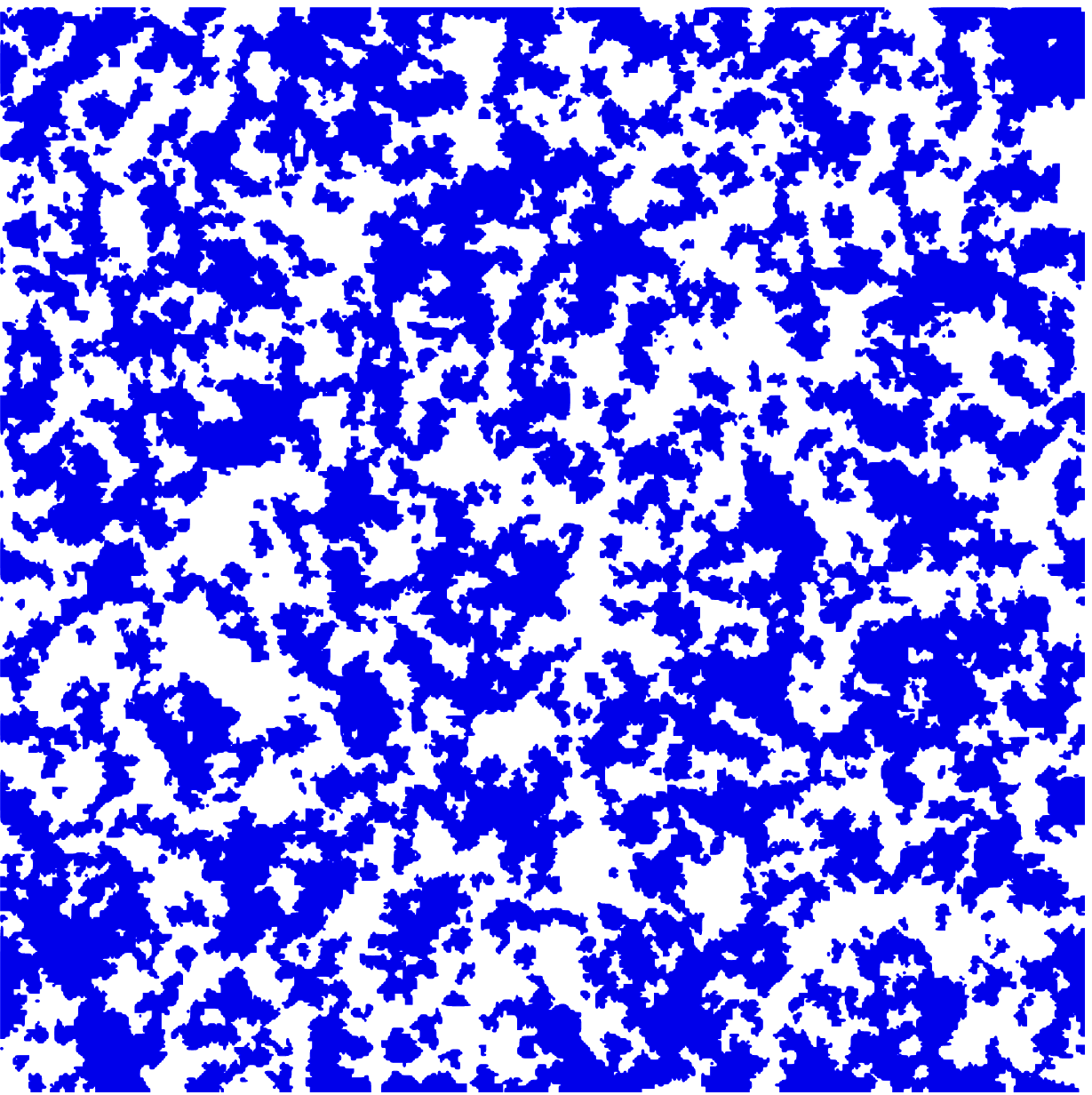}
\caption{Left panel: The preponderance of previous theoretical treatments of the effective dynamic dielectric
constant are local in nature and restricted to dispersions of well-defined dielectric scatterers (inclusions) in a matrix, as illustrated
in the leftmost panel.  In contrast to the nonlocal strong-contrast formalism presented here, earlier studies cannot treat
 the more general two-phase microstructures shown in the middle panel (spinodal decomposition pattern \cite{Ma17}) 
and rightmost panel (Debye random medium \cite{Ye98a,Ma20}), both of which
have ``phase-inversion" symmetry \cite{To02a}. Our formalism can in principle treat two-phase media of arbitrary microstructures.}
\label{examples}
\end{figure*}

Our focus in this paper is the determination  of the effective
dynamic dielectric constant tensor $\tens{\varepsilon}_e(\vect{k}_I,\omega)$  of a two-phase dielectric composite, which depends on the frequency $\omega$
or wave vector $\vect{k}_I$ of the incident radiation \cite{Sh95,Si99}. From this effective property, one
can determine the corresponding effective wave speed $c_e$ and attenuation coefficient $\gamma_e$.
The preponderance of previous homogenization theories of $\tens{\varepsilon}_e(\vect{k}_I,\omega)$
apply only in the {\it quasistatic} or long-wavelength regime \cite{Ke64b,Fr68,Ts81,Ji92}\footnote{
A notable exception is the generalized coherent potential approximation  for particle  suspensions presented in Ref. \cite{Ji92}.
However, this is not a closed-form formula for the effective dielectric constant and requires, as input, certain numerical
simulations of the electric fields.}, i.e., applicable when  $|\vect{k}_I| \ell \ll 1$,
where  $\ell$ is a characteristic heterogeneity length scale. This spectral range is the realm of nonresonant
dielectric behavior. Virtually all earlier formulas $\tens{\varepsilon}_e(\vect{k}_I,\omega)$ 
 apply to very special microstructures, namely, dielectric scatterers that are well-defined
inclusions, e.g., nonoverlapping spheres in a matrix (see Fig. \ref{examples}). Examples of such popular closed-form approximation formulas devised for spherical scatterers in a matrix 
include the Maxwell-Garnett \cite{Gar04,Ru00} and quasicrystalline \cite{Lax52,Si99,Ao02}
estimates, among others \cite{Sh95}. 

In the present investigation, we derive nonlocal homogenized constitutive relations from Maxwell's equations to obtain exact expressions for the effective dynamic dielectric constant tensor  $\tens{\varepsilon}_e(\vect{k}_I,\omega)$ of a macroscopically anisotropic two-phase medium of arbitrary microstructure that is valid well beyond the quasistatic regime, i.e., from the infinite-wavelength limit down to intermediate wavelengths ($0 \le |{\bf k}_I| \ell \lesssim 1$).
This task is accomplished by extending the strong-contrast expansion formalism, which has been used in the past exclusively for the static limit \cite{To02a} and quasistatic regime \cite{Re08a}, and establishing that the resulting homogenized constitutive relations are {\it nonlocal} in space (Sec. \ref{exact}),
i.e., the average polarization field at position $\bf x$ depends on the average electric field at other positions around $\bf x$.
(Such nonlocal relations are well known in the context
of crystal optics in order to account for ``spatial dispersion," i.e., the dependence of dielectric properties on a wave vector \cite{Ag84}.)
The terms of the strong-contrast expansion are explicitly given in terms of integrals over products of Green's functions and the $n$-point correlation functions of the random two-phase medium (defined in Sec. \ref{n-point}) to infinite order.
This representation exactly treats multiple scattering to all orders for the range of wave numbers for which our extended homogenization theory applies, i.e.,  $0 \le  |\vect{k}_I| \ell \lesssim 1$.  

It is noteworthy that the strong-contrast formalism is a significant departure from standard multiple-scattering theory \cite{Fr68,Sh95,Ts01, Ca15}, as highlighted in Sec. \ref{exact}.
Moreover, as we show there, our strong-contrast formalism has a variety of ``tuning knobs" that enable one to obtain distinctly different expansions and approximations designed for different classes of microstructures.

Because of the fast-convergence properties of strong-contrast expansions, elaborated in Sec. \ref{trunc}, their lower-order truncations yield accurate closed-form approximate formulas for the effective dielectric constant that apply for a wide class of microstructures over the aforementioned broad range of incident wavelengths, volume fractions, and contrast ratios (Sec. \ref{sec:strong-contrast}). Thus, we are able to accurately account for multiple scattering in the resonant realm (e.g., Bragg diffraction for periodic media), in contrast to the  Maxwell-Garnett and quasicrystalline approximations, which are known to break down in this spectral range. 
These nonlocal strong-contrast formulas can be regarded as approximate resummations of the expansions that still accurately capture multiple-scattering effects to all orders via the {\it nonlocal attenuation function} $F({\bf Q})$ (Sec. \ref{attenuation}).
The key quantity $F({\bf Q})$ is a functional of the spectral density $\spD{\vect{Q}}$ (Sec. \ref{2-pt}), which is straightforward to determine 
for {\it general microstructures} either theoretically, computationally or via scattering experiments.
We employ precise full-waveform simulation methods (Sec. \ref{sec:simulation}) to show that these microstructure-dependent approximations are accurate for both two-dimensional (2D) and three-dimensional (3D) ordered and disordered models of particulate composite media (Sec. \ref{sim-results}).
This validation means that they can be used to predict accurately the effective wave characteristics well beyond the quasistatic regime for a wide class of composite microstructures (Sec. \ref{predictions})  without having to perform full-blown simulations. This broad microstructure class includes
particulate media consisting of identical or polydisperse particles of general shape (ellipsoids, cubes, cylinders, polyhedra)
that may or not {\it overlap}, and cellular networks as well as media without well-defined inclusions (see Sec. \ref{trunc} for details).
Thus, our nonlocal formulas can be employed to accelerate the discovery of novel electromagnetic composites
by appropriate tailoring of the spectral densities \cite{To16b,Ch18a} and then generating the microstructures
that achieve them \cite{Ch18a}.


Although our strong-contrast formulas for the effective dynamic dielectric constant apply to periodic
two-phase media, the primary applications that we have in mind are 
{\it correlated} disordered microstructures because they can provide advantages over periodic ones with high crystallographic symmetries \cite{Lo18,Yu15} which 
include perfect isotropy and robustness against defects \cite{Fl13,Man13b}.
We are interested in  both ``garden-variety" models of disordered
two-phase media \cite{To02a} as well as exotic {\it hyperuniform} forms   \cite{To03a,Za09,To18a}.
Hyperuniform two-phase systems are characterized by an anomalous suppression of
volume-fraction fluctuations in the infinite-wavelength limit \cite{To03a,Za09,To18a}, i.e.,
the spectral density $\spD{\vect{Q}}$ obeys the condition 
\begin{equation}\label{eq:HU_condition}
\lim_{\abs{\vect{Q}}\to 0 }\spD{\vect{Q}} = 0.
\end{equation}
Such two-phase media encompass all periodic systems, many quasiperiodic media and exotic disordered ones; see Ref. \cite{To18a} and references therein.
Disordered hyperuniform systems lie between liquids and crystals; they are like liquids in that they are statistically isotropic without any Bragg peaks, and yet behave like crystals in the manner in which they suppress the large-scale density fluctuations \cite{To03a,Za09,To18a}.
Hyperuniform systems have attracted great attention over the last decade because of their deep connections to a wide range of topics that arise in physics \cite{To15,Zh16a,He17b,Ri17,Og17,Ma17,Lo18,Yu18,Wa18,To18a,Le19a,Go19,Kl20a},  materials science \cite{Ma16,Xu17,To18c,Ch18a,Ki19a}, mathematics \cite{Gh18, Br19a,To19}, and biology \cite{Ji14,To18a} as well as for their emerging technological importance in the case of the disordered varieties \cite{Fl09b,Man13b,Ma16,Le16,Fr17,Kl18,Zhang18b,To18a,Go19}.


We apply our strong-contrast formulas to predict the real and imaginary parts of the effective dielectric constant for model microstructures with typical disorder (nonhyperuniform) as well as those with exotic hyperuniform disorder (Sec. \ref{sec:models}).
We are particularly interested in exploring the dielectric properties of a special class of hyperuniform composites called disordered \textit{stealthy hyperuniform} media, which are defined to be those that possess zero-scattering intensity for a set of wave vectors around the origin \cite{Uc04b,Ba08,To15, Zh15a, Ch18a}.
        Such materials have recently been shown to be endowed with novel optical, acoustic, mechanical, and transport properties \cite{Re08a,Le16, Zh16b, De16, Gk17,Xu17, To18c,Ki20a}. Among other findings, we show that disordered hyperuniform media are generally less lossy than their nonhyperuniform counterparts.
We also demonstrate that disordered stealthy hyperuniform particulate composites exhibit singular
wave characteristics, including the capacity to act as low-pass filters that transmit waves ``isotropically” up to a selected wave number. 
They also can be engineered to exhibit refractive indices that abruptly change over a narrow range of wave numbers by tuning the spectral density.
Our results demonstrate that one can design the effective wave characteristics of a disordered composite, hyperuniform or not, by engineering the microstructure to possess tailored spatial correlations at prescribed length scales.

\section{Background}
\label{back}

\subsection{$n$-point correlation functions}
\label{n-point}

A two-phase random medium is a domain of space $\mathcal V \subseteq \mathbb{R}^d$ that is partitioned into two disjoint regions that make up $\mathcal V$: a phase 1 region $\mathcal V_1$ of volume fraction $\phi_1$ and a phase 2 region $\mathcal V_2$ of volume fraction $\phi_2$ \cite{To02a}. The phase indicator function $\mathcal I^{(i)}(\mathbf x)$ of phase $i$ for a given realization is defined as
\begin{equation} \label{indicator}
\mathcal I^{(i)}(\mathbf x)=\left \{ \begin{aligned} & 1, & \mathbf x \in \mathcal V_i,\\ & 0, & \mathbf x \notin \mathcal V_i. \end{aligned}\right.
\end{equation}
The $n$-point correlation function $S_n^{(i)}$ for phase $i$ is defined by \cite{To82b,To02a}:
\begin{align}\label{Sndef}
S_n^{(i)}(\mathbf{x}_1, \ldots, \mathbf{x}_n) = \left\langle\prod_{j=1}^n {\cal I}^{(i)}(\mathbf{x}_j)\right\rangle,
\end{align}
where angular brackets denote an ensemble average over realizations. The function $S_n^{(i)}(\mathbf{x}_1, \ldots, \mathbf{x}_n)$ has a probabilistic interpretation: It gives the probability of finding the ends of the vectors $\mathbf{x}_1, \ldots, \mathbf{x}_n$
all in phase $i$.  
For statistically homogeneous media, $S_n^{(i)}(\mathbf{x}_1, \ldots, \mathbf{x}_n)$ is translationally invariant
and, in particular, the one-point function is position independent, i.e., 
$S_1^{(i)}(\mathbf{x_1}) = \phi_i$.

\subsection{Two-point statistics}
\label{2-pt}

For statistically homogeneous media, the two-point correlation function for phase 2 is simply related to that for phase 1 via the expression $S^{(2)}_2({\bf r}) = S^{(1)}_2({\bf r}) - 2\phi_1 + 1$, and hence the \textit{autocovariance} function is given by
\begin{equation}
\label{Auto}
\chi_{_V}({\bf r}) \equiv  S^{(1)}_2({\bf r}) - {\phi_1}^2 =  S^{(2)}_2({\bf r}) - {\phi_2}^2,
\end{equation}
which we see is the same for phase 1 and phase 2. Thus, $\chi_{_V}({\bf r}=0)=\phi_1\phi_2$ and, assuming the medium possesses no long-range order, $\lim_{|{\bf r}| \rightarrow \infty} \chi_{_V}({\bf r})=0$.
 For statistically homogeneous and isotropic media, ${\chi}_{_V}({\bf r})$ depends only on the magnitude of its argument $r=|\bf r|$, and hence is a radial function. In such cases, its slope at the origin is directly related to the {\it specific surface} $s$ (interface area per unit volume); i.e., asymptotically, we have  \cite{To02a}
\begin{equation}
\chi_{_V}({\bf r})= \phi_1\phi_2 - \beta(d) s \;|{\bf r}| + {\cal O}(|{\bf r}|^2),
\label{specific}
\end{equation}
where
\begin{equation}
\beta(d)= \frac{\Gamma(d/2)}{2\sqrt{\pi} \Gamma((d+1)/2)},
\label{beta}
\end{equation}
and $\Gamma(x)$ is the gamma function.

The nonnegative spectral density ${\tilde \chi}_{_V}({\bf Q})$, which is proportional to scattering intensity \cite{De57},
is  the Fourier transform of $\chi_{_V}({\bf r})$, i.e.,
\begin{equation}
{\tilde \chi}_{_V}({\bf Q}) = \int_{\mathbb{R}^d} \chi_{_V}({\bf r}) \,e^{-i{\bf Q \cdot r}} {\rm d} {\bf r} \ge 0, \qquad \mbox{for all} \; {\bf Q},
\label{def-spec}
\end{equation}
where $\vect{Q}$ represents the momentum-transfer wave vector.
For statistically homogeneous media, the spectral density must obey the following sum rule \cite{To20a}:
\begin{equation}
\frac{1}{(2\pi)^d}\int_{\mathbb{R}^d} {\tilde \chi}_{_V}({\bf Q})\, \dd{\bf Q}= \chi_{_V}({\bf r}=0)=\phi_1\phi_2.
\label{sum}
\end{equation}
For isotropic media, the spectral density depends only on $Q=|{\bf Q}|$ and, as a consequence of Eq. (\ref{specific}), its large-$k$ 
behavior is controlled by the following power-law form:
\begin{equation}
{\tilde \chi}_{_V}({\bf Q}) \sim \frac{\gamma(d)\,s}{Q^{d+1}}, \qquad Q \rightarrow \infty,
\label{decay}
\end{equation}
where
\begin{equation}
\gamma(d)=2^d\,\pi^{(d-1)/2} \,\Gamma((d+1)/2) \,\beta(d)
\end{equation}
is a $d$-dimensional  constant and $\beta(d)$ is given by Eq. (\ref{beta}).

\subsection{Packings}

We call a {\it packing} in $\mathbb{R}^d$ a collection of nonoverlapping particles
\cite{To18b}. In the case of a packing of identical spheres of radius $a$ at number density $\rho$,
the spectral density $\tilde{\chi}_{_V}(\textbf{Q})$ is directly related to the
structure factor $S({\bf Q})$ of the sphere centers \citep{To02a,To16b}:
\begin{align}
  \tilde{\chi}_{_V}(\textbf{Q})= \phi_2 \, \tilde{\alpha}_2(Q;a)\, S(\textbf{Q}),
  \label{chi-packing}
\end{align}
where  
\begin{align}
  \tilde{\alpha}_2(Q;a) = \frac{1}{v_1(a)}\left( \frac{2\pi a}{Q} \right)^{d}\, J^2_{d/2}(Qa),
  \label{chi-large}
\end{align}
$\fn{J_\nu}{x}$ is the Bessel function of the first kind of order $\nu$, $\phi_2=\rho v_1(a)$ is the {\it packing
fraction} (fraction of space covered by the spheres), and
\begin{equation}
v_1(a)=\frac{\pi^{d/2} a^d}{\Gamma(1+d/2)}
\label{v1}
\end{equation}
is the $d$-dimensional volume of a sphere of radius $a$.

\subsection{Hyperuniformity and volume-fraction fluctuations}
\label{hyper}

Originally introduced in the context of point configurations \cite{To03a}, the hyperuniformity concept was generalized to treat two-phase media \cite{Za09}.
 Here the phase volume fraction fluctuates within a spherical window of radius $R$, which can be characterized by the local volume-fraction variance $\sigma_{_V}^2(R)$.
This variance is directly related to integrals involving either the autocovariance function $\chi_{_V}({\bf r})$ \cite{Lu90b} or the spectral density $\tilde{\chi}_{_V}({\bf Q})$ \cite{Za09}.

For typical disordered two-phase media, the variance  $\sigma_{_V}^2(R)$ for large $R$ goes to zero like $R^{-d}$ \cite{Lu90b,Qu97b,To02a}.
However, for hyperuniform disordered media, $\sigma_{_V}^2(R)$  goes to zero asymptotically more rapidly than the inverse of the window volume, i.e., faster than $R^{-d}$, which is equivalent to the condition \eqref{eq:HU_condition} on the spectral density.
{\it Stealthy hyperuniform} two-phase media  are a subclass of hyperuniform systems in which $\tilde{\chi}_{_V}(\mathbf{Q})$ is zero for a range of wave vectors around the origin, i.e.,
\begin{equation}
\tilde{\chi}_{_V}(\mathbf{Q})= 0 \qquad \mbox{for}\; 0 \le |{\bf Q}| \le Q_\text{U},
\label{stealthy}
\end{equation}
where $Q_\text{U}$ is some positive number. 

As in the case of hyperuniform point configurations \cite{To03a,Za09,Za11b,To18a},  there are three different scaling regimes (classes) 
that describe the associated large-$R$ behaviors of the volume-fraction variance when the spectral density goes to zero as a power law  ${\tilde \chi}_{_V}({\bf Q})\sim |{\bf Q}|^\alpha$ as $|\bf Q|\to 0$:
\begin{align}  
\sigma^2_{_V}(R) \sim 
\begin{cases}
R^{-(d+1)}, \quad\quad\quad \alpha >1 \qquad &\text{(Class I)},\\
R^{-(d+1)} \ln R, \quad \alpha = 1 \qquad &\text{(Class II)},\\
R^{-(d+\alpha)}, \quad 0 < \alpha < 1\qquad  &\text{(Class III)},
\end{cases}
\label{sigma-V-asy}
\end{align}
where the exponent $\alpha$ is a positive constant. 
Thus, the characteristics length of the representative elementary volume
for a hyperuniform medium will depend on the hyperuniformity class (scaling).
Class I is the strongest form of hyperuniformity, which includes all perfect periodic packings as well as some disordered packings,  such as disordered stealthy packings described in Sec. \ref{stealth}. 

\subsection{Popular effective-medium approximations}
\label{approx}

Here we explicitly state the specific functional forms of an extended Maxwell-Garnett approximation
and quasicrystalline approximation for the effective dynamic dielectric constant $\varepsilon_e(k_1)$ 
of isotropic media composed of identical spheres of dielectric constant $\varepsilon_2$ embedded in a matrix phase of dielectric constant $\varepsilon_1$.
In Sec. \ref{sim-results}, we compare the predictions of these formulas to those of our nonlocal approximations.
The small-wave-number expansions of these popular approximations are provided in the Supplementary Material \cite{SM}. There we also provide the corresponding asymptotic behaviors of our strong-contrast
approximations.

\subsubsection{Maxwell-Garnett approximation}

Maxwell-Garnett approximations (MGAs) \cite{Gar04,Ru00} are derived by substituting the dielectric polarizability of a single dielectric sphere into the Clausius-Mossotti equation \cite{Ru00}, which consequently ignores the spatial correlations between the particles. 
In three dimensions, we utilize the following extended MGA that makes use of the exact electric dipole polarizability $\fn{\alpha_e}{k_1}$ of a single dielectric sphere of radius $a$ \cite{Do89}:
\begin{align} \label{eq:Maxwell-Garnett_3D}
\frac{\fn{\varepsilon_e}{k_1} - \varepsilon_1}{\fn{\varepsilon_e}{k_1} +2 \varepsilon_1} 
    =&
    \phi_2  \fn{\alpha_e}{k_1} /a^3,
\end{align}
where 
$$
\fn{\alpha_e}{k_1} = 
\frac{3i}{2 {k_1}^3}
\frac{
    m \fn{\psi_1}{m k_1 a}\fn{\psi_1 '}{k_1 a} - \fn{\psi_1}{k_1 a} \fn{\psi_1 ' }{m k_1 a}  
}{
    m \fn{\psi_1}{m k_1 a}\fn{\xi_1 '}{k_1 a} - \fn{\xi_1}{k_1 a} \fn{\psi_1 ' }{m k_1 a}
},
$$
$m\equiv \sqrt{\varepsilon_2 / \varepsilon_1}$, $\fn{\psi_1}{x} \equiv x \fn{j_1}{x}$, $\fn{\xi_1}{x} \equiv x \fn{h_1 ^{(1)}}{x} $, the prime symbol $(')$ denotes the derivative of a function,  $\fn{h_1 ^{(1)}}{x} $ is the spherical Hankel function of the first kind of order 1, and $\fn{j_1}{x}$ is the spherical Bessel function of the first kind of order 1.

The 2D analog of Eq. \eqref{eq:Maxwell-Garnett_3D} can be obtained by using the dynamic dielectric polarizability $\alpha_e$ of a dielectric cylinder of radius $a$ given 
in Ref. \cite{Si07}:
\begin{align} \label{eq:Maxwell-Garnett_2D}
\frac{\fn{\varepsilon_e}{k_1} - \varepsilon_1}{\fn{\varepsilon_e}{k_1} + \varepsilon_1} 
    =&
    \frac{\phi_2}{2\pi}\fn{\alpha_e}{k_1}/a^2  ,
\end{align}
where
\begin{align*}
\fn{\alpha_e}{k_1} =& 
\frac{4 (\varepsilon_2 - \varepsilon_1)}{i {k_1}^2 m \varepsilon_1} \fn{J_1}{m k_1 a} \big[ \fn{J_1 '}{m k_1 a} \Hankel{1}{k_1 a} \\
&- m \fn{J_1}{m k_1 a}
\fn{{\mathcal{H}_1^{(1)}}'}{k_1 a}\big]^{-1},
\end{align*}
where $\Hankel{\nu}{x}$ is the Hankel function of the first kind of order $\nu$.
Here, only transverse-electric (TE) polarization is considered; i.e., the electric field is  perpendicular to the axis of the cylinder.

As with all MGA theories, formulas \eqref{eq:Maxwell-Garnett_3D} and \eqref{eq:Maxwell-Garnett_2D} neglect spatial correlations between the  particles and hence are only valid for low inclusion packing fractions. 
In the static limit, Eqs. \eqref{eq:Maxwell-Garnett_3D} and \eqref{eq:Maxwell-Garnett_2D} reduce to
the Hashin-Shtrikman estimates $\varepsilon_\mathrm{HS}$; see relation \eqref{eq:HS-bound}.

\subsubsection{Quasicrystalline approximations}

The quasicrystalline approximation (QCA) for the quasistatic effective dynamic dielectric constant $\fn{\varepsilon_e}{k_1}$ employs the ``effective" Green's function of spherical scatterers up to the level of the pair correlation function $\fn{g_2}{\vect{r}}$ \cite{Lax52,Si99,Ao02}.
However, the QCA accounts only for the structure factor in the infinite-wavelength limit \cite{Ao02} [i.e., $\fn{S}{0} = 1+\rho \int(\fn{g_2}{\vect{r}}-1)\dd{\vect{r}}$] and consequently, spatial correlations at finite wavelengths are ignored.  
The QCA for $d=3$ can be explicitly written as follows  \cite{Si99}: 
\begin{align}
&{\phi_2}^2\beta_{21}\qty[\frac{\fn{\varepsilon_e}{k_1} - \varepsilon_1}{\fn{\varepsilon_e}{k_1} +2 \varepsilon_1}]^{-1} = \phi_2 - i \Bigg\{\frac{2}{3}\phi_2 \fn{S}{0} (k_1 a)^3 \nonumber \\
&\times\qty[1+i \frac{2}{3(1-\beta_{21} \phi_2)} (k_1 a)^3 \fn{S}{0}]^{-1}\Bigg\}\beta_{21},\label{eq:quasicrystalline_low-con2}
\end{align}
where $\beta_{21}$ is defined in Eq. \eqref{eq:polarizability}.
Interestingly, the QCA predicts that the hyperuniform composites [$S(0)=0$] will be transparent for all wave numbers, which cannot be true for stealthy hyperuniform media [cf. Eq. \eqref{stealthy}], since the transparency interval must be finite; see Sec. \ref{sim-results}.
Note that Eq. (\ref{eq:quasicrystalline_low-con2}) is the complex conjugate of the one given in Ref. \cite{Si99} so that it is consistent with the sign convention for the imaginary part $\Im[\varepsilon_e]$ 
used here.

\section{Exact Strong-Contrast Expansions and Nonlocality}
\label{exact}

The original strong-contrast expansions for the effective dynamic dielectric constant obtained by Rechtsman and Torquato \cite{Re08a} were derived from homogenized constitutive relations that are local in space and hence are strictly valid only in the quasistatic regime. In the present work, we follow the general strong-contrast formalism of Torquato \cite{To02a} that was devised for the purely static problem \cite{To02a} and show that it naturally leads to exact homogenized constitutive relations for the averaged fields that are nonlocal in space.
The crucial consequences of this development are exact expressions for the effective dynamic dielectric constant tensor $\tens{\varepsilon}_e(\vect{k}_I)$ for a macroscopically anisotropic medium of arbitrary microstructure into which a plane wave of wave vector $\vect{k}_I$ is incident. 
These expressions for $\tens{\varepsilon}_e(\vect{k}_I)$ are valid from the infinite-wavelength limit down to wavelengths ($\lambda = 2\pi /\abs{\vect{k}_I}$) on the order of the heterogeneity length scale $\ell$. 
We then briefly explain how our theory departs substantially from standard multiple-scattering theory \cite{Fr68,Sh95,Ts01,Ca15}. 
We explicitly show they necessarily require complete microstructural information, as embodied in the infinite set of $n$-point correlation functions (Sec. \ref{n-point}) of the composite.
We also describe the fast-convergence properties of strong-contrast expansions and their consequences for extracting accurate approximations for $\tens{\varepsilon}_e(\vect{k}_I)$.

\subsection{Strong-contrast expansions}

\begin{figure}[h]
\includegraphics[width = 0.5\textwidth]{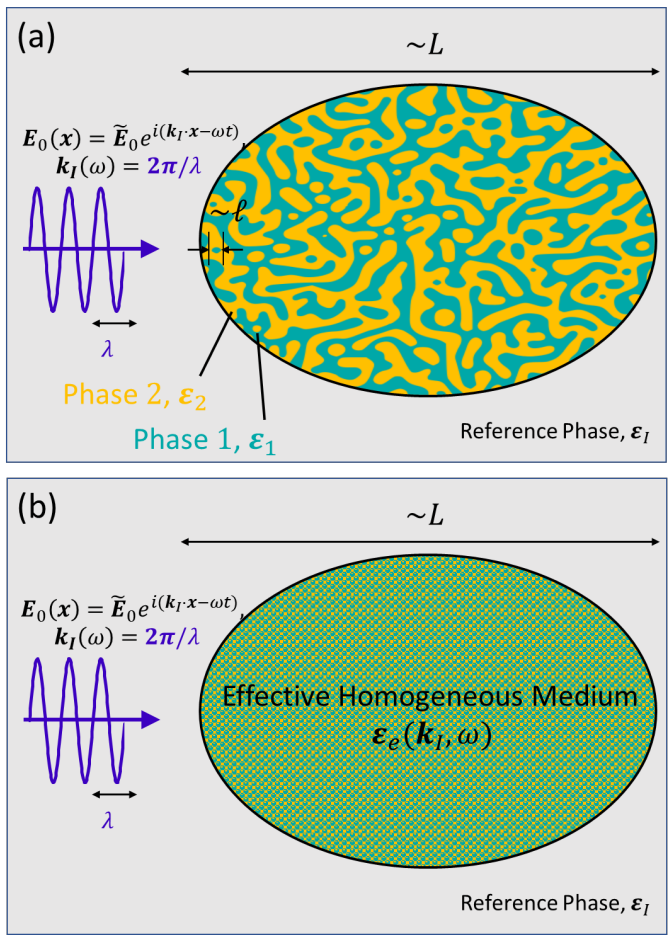}
\caption{(a) Schematic of a large $d$-dimensional ellipsoidal, macroscopically anisotropic two-phase composite medium embedded in an infinite reference phase of dielectric constant tensor $\tens{\varepsilon}_I$ (gray regions) under an applied electric field $\fn{\vect{E}_0}{\vect{x}} = \tilde{\vect{E}}_0 \fn{\exp}{
    i(\vect{k}_I\cdot\vect{x}-\omega t)}$ of a frequency $\omega$ and a wave vector $\vect{k}_I$ at infinity.
The wavelength $\lambda$ associated with the applied field can span from the quasistatic regime ($2\pi \ell/\lambda  \ll 1$) down to the intermediate-wavelength regime ($2\pi \ell/\lambda \lesssim 1$), where $\ell$ is a characteristic heterogeneity length scale.  
(b) After homogenization, the same ellipsoid can be regarded as a homogeneous specimen with an effective dielectric constant $\tens{\varepsilon_e}({\vect{k}_I,\omega})$, which depends on $\omega$ and $\vect{k}_I$. As noted in the main text, we omit the $\omega$ dependence of $\tens{\varepsilon_e}$
because (without loss of generally) we assume a linear dispersion relation between $|\vect{k}_I|$ and $\omega$.
\label{specimen}}
\end{figure}

Consider a macroscopically large ellipsoidal two-phase statistically homogeneous but anisotropic composite specimen in $\R^d$ embedded inside an infinitely large reference phase $I$ with a dielectric constant tensor $\tens{\varepsilon}_I$.
The microstructure is perfectly general, and it is assumed that a characteristic heterogeneity length scale $\ell$ is much smaller than the specimen size, i.e., $\ell  \ll L$.
The shape of this specimen is purposely chosen to be nonspherical since any rigorously correct expression for the effective property must ultimately be independent of the shape of the composite specimen in the infinite-volume limit.
It is  assumed that the applied or incident electric field $\fn{\vect{E}_0}{\vect{x}}$ is a  plane wave of an angular frequency $\omega$ and wave vector $\vect{k}_I$ in the reference phase, i.e.,
\begin{equation}
\fn{\vect{E}_0}{\vect{x}} = \tilde{\vect{E}}_0\fn{\exp}{i(\vect{k}_I\cdot \vect{x} - \omega t)}.
\label{applied}
\end{equation} 
Our interest is the exact expression for the effective dynamic dielectric constant  tensor $\tens{\varepsilon}_e({\bf k}_I,\omega)$.  
Without loss of generality, we assume a linear dispersion relation in the reference phase, i.e., $k_I\equiv \abs{\vect{k}_I} = \sqrt{\varepsilon_I}\omega /c$, 
where $c$ is the speed of light in vacuum, and thus we henceforth do not explicitly indicate the dependence of functions  on $\omega$. 
The composite is assumed to be nonmagnetic, implying that the phase magnetic permeabilities are identical, i.e., $\mu_1 = \mu_2=\mu_0$, where $\mu_0$ is the magnetic permeability of the vacuum.
For simplicity, we assume real-valued, frequency-independent isotropic phase dielectric constants  $\varepsilon_1$ and $\varepsilon_2$.
Nonetheless, the composite can be generally lossy (i.e., $\tens\varepsilon_e$ is complex-valued) due to scattering from the inhomogeneities in the local dielectric constant.
It is noteworthy that our results can be straightforwardly extended to phase dielectric constants that are complex-valued (dissipative media), but this is not done in the present work.

Here we present a compact derivation of strong-contrast expansions. It follows the general formalism of Torquato \cite{To02a} closely but departs from it at certain key steps when establishing the nonlocality of the homogenized constitutive relation.
(A detailed derivation is given in the Supplementary Material \cite{SM}.)
For simplicity, we take the reference phase $I$ to be phase $q$ (equal to 1 or 2).
Under the aforementioned assumptions, the local electric field $\fn{\vect{E}}{\vect{x}}$ solves the time-harmonic Maxwell equation \cite{Re08a}:
\begin{equation}
    \curl{\curl{\fn{\vect{E}}{\vect{x}}}} - {k_q}^2 \fn{\vect{E}}{\vect{x}}= \qty(\frac{\omega}{c})^2 \fn{\vect{P}}{\vect{x}}  , \label{eq:wave-equation_isotropic_reference}
    \end{equation}
where  $\fn{\vect{P}}{\vect{x}}$ is the {\it polarization field} given by    
    \begin{equation}
    \fn{\vect{P}}{\vect{x}} \equiv    \qty[\fn{{\varepsilon}}{\vect{x}} - {\varepsilon}_q] \fn{\vect{E}}{\vect{x}} \label{eq:def_polarization}
    \end{equation}
and 
\begin{equation}
\varepsilon({\bf x})= (\varepsilon_p -\varepsilon_q){\cal I}^{(p)}({\bf x}) +\varepsilon_q
\end{equation}
is the local dielectric constant, and ${\cal I}^{(p)}({\bf x})$ is the indicator function for phase $p$ [cf. Eq. (\ref{indicator})]. 
The vector $ \fn{\vect{P}}{\vect{x}}$ is the induced flux field  relative to reference phase $q$ due to the presence of phase $p$, and hence is zero in the reference phase $q$ and nonzero in the ``polarized" phase $p$  ($p\neq q$).

Using the Green's function formalism, the local electric field can be expressed in terms of the following
integral equation \cite{Re08a,To02a}:
    \begin{align}
    \fn{\vect{E}}{\vect{x}}  = &  \fn{\vect{E}_0}{\vect{x}} + \int \fn{\tens{G}^{(q)}}{\vect{\vect{x}-\vect{x}'}} \cdot \fn{\vect{P}}{\vect{x}'} \dd{\vect{x}'} ,
    \label{eq:GreenMethod}
    \end{align}    
where the second-rank tensor Green's function $\fn{\tens{G}^{(q)}}{\vect{r}}$ associated with the reference phase $q$ is given by \footnote{
The Green's function in this work (\ref{eq:relation_GreenFn_D_H}) differs from the one given in Ref. \cite{Re08a} by a
multiplicative factor $(\omega/c)^2$. Thus, Eq. \eqref{eq:relation_GreenFn_D_H} converges to its static counterpart in the static
limit (i.e., $\omega \to 0$).}
\begin{equation}
    \fn{\tens{G}^{(q)}}{\vect{r}} = -\tens{D}^{(q)}\fn{\delta}{\vect{r}} + \fn{\tens{H}^{(q)}}{\vect{r}}, 
\label{eq:relation_GreenFn_D_H}
\end{equation}
$\tens{D}^{(q)}$ is a constant  second-rank tensor  that arises when one excludes an infinitesimal region around the position of the singularity $\vect{x}'=\vect{x}$ in the Green's function, and $\fn{\tens{H}^{(q)}}{\vect{r}}$ represents the contribution outside of the ``exclusion" region:
\begin{align} 
  \fn{\ten{H^{(q)}}{ij}}{\vect{r}} =&
    i\frac{\pi}{2  \varepsilon_q}\qty(\frac{k_q}{2\pi r})^{d/2} \big\{ \qty[k_q r \Hankel{d/2-1}{k_qr} -\Hankel{d/2}{k_qr}]\ten{\delta}{ij} 
    \nonumber \\
    & +k_qr \Hankel{d/2+1}{k_q r} \uvect{r}_i \uvect{r}_j\big\},\label{eq:H-tensor-supp}
\end{align}
where $\uvect{r} \equiv \vect{r} / \abs{\vect{r}}$ is a unit vector directed to $\vect{r}$, and $\Hankel{\nu}{x}$ is the Hankel function of the first kind of order $\nu$.
The Fourier transform of  Eq. \eqref{eq:relation_GreenFn_D_H} is particularly simple and concise:
\begin{equation}
    \fn{\ten{\tilde{G}^{(q)}}{ij} }{\vect{k}} = \frac{1}{\varepsilon_q} \frac{{k_q}^2 \delta_{ij} -k_i k_j}{k^2 - {k_q}^2}. \label{eq:FT_GreenFn}
    \end{equation}    
Note that Eq. \eqref{eq:FT_GreenFn} is independent of the shape of the exclusion region, which stands in contrast to the shape-dependent  Fourier transform of $\fn{\ten{H^{(q)}}{ij}}{\vect{r}}$; see Supplementary Material \cite{SM} for details.

The use of Eq. (\ref{eq:def_polarization}) and (\ref{eq:GreenMethod}) leads to an integral equation for the \textit{generalized cavity intensity field} $\fn{\vect{F}}{\vect{x}}$:
\begin{align}
    \fn{\vect{F}}{\vect{x}}  = &  \fn{\vect{E}_0}{\vect{x}} + \int_\epsilon \fn{\tens{H}^{(q)}}{\vect{\vect{x}-\vect{x}'}} \cdot \fn{\vect{P}}{\vect{x}'} \dd{\vect{x}'} ,
    \label{eq:GreenMethod-2}
    \end{align}
where the integral subscript $\epsilon$ indicates that the integral is to be carried out by omitting the exclusion region and then allowing it to uniformly shrink to zero. 
Here $\fn{\vect{F}}{\vect{x}}$ is related directly to $\fn{\vect{E}}{\vect{x}}$ via 
\begin{equation}
\fn{\vect{F}}{\vect{x}}= \qty{\tens{I} + \tens{D}^{(q)}[\fn{\varepsilon}{\vect{x}} - \varepsilon_q]}\cdot \vect{E}(\vect{x})  \label{eq:linearRel_F}.
\end{equation}
Using the definitions \eqref{eq:def_polarization} and \eqref{eq:linearRel_F}, we obtain a linear constitutive relation between $\fn{\vect{P}}{\vect{x}}$ and $\fn{\vect{F}}{\vect{x}}$: 
    \begin{equation}
    \fn{\vect{P}}{\vect{x}} = \fn{\tens{L}^{(q)}}{\vect{x}} \cdot \fn{\vect{F}}{\vect{x}}, \label{eq:relation_P_F}
    \end{equation}
    where 
\begin{equation}
\fn{\tens{L}^{(q)}}{\vect{x}} =  [\fn{\varepsilon}{\vect{x}} - \varepsilon_q] 
    \cdot\qty{\tens{I} + \tens{D}^{(q)}[\fn{\varepsilon}{\vect{x}} - \varepsilon_q]}^{-1}.
\label{eq:L-tensor-local}
\end{equation}

It is noteworthy that one is free to choose any convenient exclusion-region shape, provided that its boundary is sufficiently smooth.
The choice of the exclusion-region shape is crucially important because it determines the type of {\it modified} electric field that results as well as the corresponding expansion parameter in the series expansion for the effective dynamic dielectric constant tensor. 
For example, when a spheroidal-shaped exclusion region at a position $\vect{x}$ in $\R^d$ is chosen to be aligned with the polarization vector $\fn{\vect{P}}{\vect{x}}$ at that position, we have 
        \begin{equation}
        \tens{D}^{(q)} = \frac{A^*}{\varepsilon_q} \tens{I},
\label{eq:D-tensor}
        \end{equation}
        where $\tens{I}$ is the second-rank identity tensor and
$A^*\in[0,1]$ is the depolarization factor for a spheroid \cite{To02a}
with the aforementioned alignment.
In the special cases of a sphere, disk-like limit, and needle-like limit,
$A^*=1/d, 1, 0$, respectively. 
Thus, for these three cases,  Eq. \eqref{eq:L-tensor-local} yields the following shape-dependent tensor:
\begin{align}
&\fn{\tens{L}^{(q)}}{\vect{x};A^*} = \fn{\mathcal{I}^{(p)}}{\vect{x}}\tens{I} \nonumber \\
\quad &\times 
    \begin{cases}
    d\varepsilon_q \beta_{pq}, &A^* = 1/d~(\text{spherical})\\
    \varepsilon_q\qty(1 - \varepsilon_q/\varepsilon_p), &A^* = 1~(\text{disk-like})\\
    \varepsilon_q\qty(\varepsilon_p/\varepsilon_q - 1), &A^* = 0~(\text{needle-like})
    \end{cases},\label{eq:exclusion-region}
\end{align}
    where $\beta_{pq}$ is the dielectric \textit{polarizability} defined by
    \begin{equation}     \label{eq:polarizability}
\beta_{pq} \equiv \frac{\varepsilon_p - \varepsilon_q}{\varepsilon_p +
(d-1)\varepsilon_q}.
\end{equation}
Moreover, in these three cases, the generalized cavity intensity field $\fn{\vect{F}}{\vect{x};A^*}$ reduces to
\begin{equation}
\fn{\vect{F}}{\vect{x};A^*} \to 
\begin{cases}
\fn{\vect{E}}{\vect{x}}+ \frac{\displaystyle \fn{\vect{P}}{\vect{x}}}{\displaystyle d\varepsilon_q}, &A^* = 1/d~(\text{spherical})\\
\frac{\displaystyle \fn{\vect{D}}{\vect{x}}}{\displaystyle \varepsilon_q}, & A^* = 1~(\text{disk-like})\\
\fn{\vect{E}}{\vect{x}}, & A^* = 0~(\text{needle-like}),
\end{cases}
\end{equation}
respectively, where $\fn{\vect{D}}{\vect{x}}$ is the  displacement field.
Importantly, among these three cases, the original and modified strong-contrast expansions arise only when the exclusion region is taken to be a sphere, as we elaborate below.

We now show that our formalism yields an exact relation between the polarization field $\vect{P}(\vect{x})$ and the applied field $\vect{E}_0(\vect{x})$ that is nonlocal in space.
It is more convenient at this stage to utilize a compact linear operator notation, which enables us to express the integral equation \eqref{eq:GreenMethod-2} as 
\begin{align}
\vect{F}     = \vect{E}_0 + \tens{H} \vect{P} \label{eq:integralEq_F},
\end{align}
where we temporarily drop the superscript $q$.
A combination of this equation with Eq. (\ref{eq:relation_P_F}) yields the following integral equation for the polarization field
\begin{equation}
        \vect{P} = \tens{L}\vect{E}_0 + \tens{L}\tens{H}\vect{P}.
\label{P-P}
\end{equation}
The desired nonlocal relation is obtained from Eq. (\ref{P-P}) by successive substitutions: 
\begin{align}
\vect{P} =  \tens{S}\vect{E}_0. \label{eq:S-tensor}
\end{align}
where
\begin{equation}
\tens{S}=[\tens{I} - \tens{L}\tens{H}]^{-1}\tens{L} ,
\label{SO}
\end{equation}
is a {\it generalized scattering operator} that has superior mathematical properties compared to the
scattering operator $\tens{\mathcal{T}}$ that arises in standard 
multiple-scattering theory \cite{Fr68, Sh95, Ts01}, as we elaborate below in Remark \ref{remark8}.
More explicitly, the nonlocal relation \eqref{eq:S-tensor} can be expressed as
\begin{align}
\fn{\vect{P}}{\vect{1}} = & \int_\epsilon \dd{\vect{2}} \fn{\tens{S}}{\vect{1},\vect{2}} \cdot \fn{\vect{E}_0}{\vect{2}}, \label{eq:P2E0}        
\end{align}
where boldface numbers $\vect{1},\vect{2}$ are shorthand notations for position vectors $\vect{r}_1,\vect{r}_2$.
Ensemble averaging Eq. (\ref{eq:P2E0}) and invoking statistical homogeneity yields the convolution relation
\begin{equation}
\fn{\E{\vect{P}}}{\vect{1}} 
= \int_\epsilon \dd{\vect{2}} \fn{\E{\tens{S}}}{\vect{1}-\vect{2}}\cdot \fn{\vect{E}_0}{\vect{2}},\label{eq:ensemble-average_Polarization}
\end{equation}
where  the operator $\E{\tens{S}}$ depends on relative positions, i.e., $\fn{\E{\tens{S}}}{\vect{1},\vect{2}} = \fn{\E{\tens{S}}}{\vect{1}-\vect{2}}$, and angular brackets denote an ensemble average.
Formally, the nonlocal relation (\ref{eq:ensemble-average_Polarization}) is the same as the one given in Torquato \cite{To02a} for the static problem, but nonlocality was not explicitly invoked there. 
Taking the Fourier transform of (\ref{eq:ensemble-average_Polarization}) yields a compact Fourier representation of this nonlocal relation, namely,
\begin{equation}
\fn{\widetilde{\E{\vect{P}}}}{\vect{k}} 
= \fn{\widetilde{\E{\tens{S}}}}{\vect{k}}\cdot \fn{\tilde{\vect{E}}_0}{\vect{k}}, \label{eq:ensemble-average_Polarization-Fourier}
\end{equation}
where $\fn{\widetilde{\E{f}}}{\vect{k}} \equiv \int \fn{\E{f}}{\vect{x}}\fn{\exp}{-i\vect{k} \cdot \vect{x}} \dd{\vect{x}}$.
From Eq. \eqref{applied}, $\fn{\tilde{\vect{E}}_0}{\vect{k}} = \tilde{\vect{E}}_0 \fn{\delta}{\vect{k}-\vect{k}_q}$, implying that the wave vector $\vect{k}$ in Eq. \eqref{eq:ensemble-average_Polarization-Fourier} must be identical to $\vect{k}_q$.

As in the static case \cite{Se89,To02a} and quasistatic regime \cite{Re08a}, the ensemble-averaged operator $\fn{\E{\tens{S}}}{\vect{r}}$, which is given explicitly in the Supplementary Material \cite{SM} in terms of the $n$-point correlation functions and  products of the tensor $\tens{H}(\vect{r})$,  depends on the shape of the macroscopic ellipsoidal composite specimen (see Fig. \ref{specimen}).
This shape-dependence arises from because $\tens{H}$ decays like $r^{-d}$ for large $r$, and hence, $\fn{\E{\tens{S}}}{\vect{r}}$ involves conditionally convergent integrals \cite{To02a}.
To avoid such conditional convergence issues, we follow previous strong-contrast formulations by seeking to eliminate
the applied field $\vect{E}_0$ in Eq. (\ref{eq:ensemble-average_Polarization}) in favor of the average cavity field  $\fn{\E{\vect{F}}}{\vect{r}}$ in order to get a corresponding nonlocal homogenized constitutive relation between  $\fn{\E{\vect{P}}}{\vect{r}}$ and  $\fn{\E{\vect{F}}}{\vect{r}}$ or vice versa.
Thus, solving for $\vect{E}_0$ in Eq. (\ref{eq:ensemble-average_Polarization}) and substituting into the ensemble average of (\ref{eq:integralEq_F}) yields 
\begin{equation}
\langle \vect{F}\rangle =[\E{\tens{S}}^{-1}+ \tens{H}] \E{\vect{P}}.
\label{F-P}
\end{equation}
\noindent Inverting this expression leads to the following {\it nonlocal} constitutive relation:
\begin{equation}
\fn{\E{\vect{P}}}{\vect{1}} 
= \int \dd{\vect{2}} \tens{L}_e^{(q)}(\vect{1}-\vect{2}) \cdot  \fn{\E{\vect{F}}}{\vect{2}},\label{P-F}
\end{equation}
where $ \tens{L}_e^{(q)}(\vect{r})$ is a kernel that is derived immediately below and explicitly given by 
\begin{align}
    \fn{\tens{L}_e^{(q)}}{\vect{r}} \equiv& 
    \int \dd{\vect{r}'} \qty[\fn{\tens{\varepsilon}_e}{\vect{r}'} - \varepsilon_q \tens{I}\fn{\delta}{\vect{r}'}] \cdot \Big[\tens{I}\fn{\delta}{\vect{r}-\vect{r}'} 
    \nonumber \\
    &+ \tens{D}^{(q)}\cdot \qty(\fn{\tens{\varepsilon}_e}{\vect{r}-\vect{r}'} - \varepsilon_q \tens{I}\fn{\delta}{\vect{r}-\vect{r}'})\Big]^{-1}.
\label{kernel}
\end{align}
We are not aware of any previous work that derives such an exact nonlocal homogenized constitutive relation (\ref{P-F}) from first principles.

Note that the support $\ell_s$ of the kernel $ \tens{L}_e^{(q)}(\vect{r})$ relative to the incident wavelength $\lambda$ determines the degree of spatial dispersion.
When $\lambda$ is finite, the relation between $\fn{\E{\vect{P}}}{\vect{x}}$ and $\fn{\E{\vect{F}}}{\vect{x}}$ in Eq. (\ref{P-F}) is nonlocal in space.
In the regime $\ell_s \ll \lambda$, the nonlocal relation (\ref{P-F})
can be well approximated by the local relation $\fn{\E{\vect{P}}}{\vect{x}} \approx \qty[\int \fn{\tens{L}_e^{(q)}}{\vect{x}'}\dd{\vect{x}'}] \cdot \fn{\E{\vect{F}}}{\vect{x}}$.
Indeed, in the static limit, $ \tens{L}_e^{(q)}(\vect{r})$ tends to a Dirac delta function $\fn{\delta}{\vect{r}}$, expression (\ref{P-F}) becomes the {\it position-independent} local relation
\begin{equation}
\E{\vect{P}}
=  \tens{L}_e^{(q)} \cdot \E{\vect{F}} \label{P-F-2}
\end{equation}
derived earlier \cite{To02a}.

The nonlocal constitutive relation in direct space, Eq. (\ref{P-F}), can be reduced to a linear product form in Fourier space by taking  the Fourier transform of Eq. (\ref{P-F}):
\begin{align}
\fn{\widetilde{\E{\vect{P}}}}{\vect{k}_q} = \fn{\tens{L}_e^{(q)}}{\vect{k}_q}\cdot \fn{\widetilde{\E{\vect{F}}}}{\vect{k}_q}.
    \label{eq:Le_tensor}
    \end{align}
The wave-vector-dependent effective tensor  $\fn{\tens{L}_e^{(q)}}{\vect{k}_q}$ is postulated (see discussion in the Supplementary Material \cite{SM}) to be given by
    \begin{equation}    \label{eq:def_Le}
    \fn{\tens{L}_e ^{(q)}}{\vect{k}_q}\equiv
    [\fn{\tens{\varepsilon}_e}{\vect{k}_q} - \varepsilon_q \tens{I}]
    \cdot \qty{\tens{I} + \tens{D}^{(q)}\cdot [\fn{\tens{\varepsilon}_e}{\vect{k}_q} - \varepsilon_q \tens{I}]
}^{-1}.
    \end{equation}
This linear  fractional form for $\fn{\tens{L}_e^{(q)}}{\vect{k}}$ is consistent with the one derived for the static limit \cite{To02a} and for the quasistatic regime  \cite{Re08a}.
Taking the inverse Fourier transform of Eq. (\ref{eq:def_Le}) yields its corresponding direct-space representation Eq. (\ref{kernel}). 
Taking the Fourier transform of Eq. (\ref{F-P}) yields
\begin{align}
\fn{\widetilde{\E{\vect{F}}}}{\vect{k}_q}  
    =&
    \qty[{\fn{\widetilde{\E{\tens{S}}}}{\vect{k}_q} }^{-1} + \fn{\tens{\tilde{H}}}{\vect{k}_q} ] \cdot \fn{\widetilde{\E{\vect{P}}}}{\vect{k}_q} .
\label{final}
\end{align}

\begin{widetext}
Comparing Eq. (\ref{eq:Le_tensor}) to Eq. (\ref{final}), and specifically choosing a spherical exclusion region, as discussed in Eq. \eqref{eq:exclusion-region},  yields the desired exact strong-contrast expansions for general macroscopically anisotropic two-phase media:
\begin{align}
{\phi_p}^2 \beta^2_{pq}
[\tens{\varepsilon}_e(\vect{k}_q) +(d-1)
\varepsilon_q\tens{I}]\cdot [\tens{\varepsilon}_e(\vect{k}_q) -
\varepsilon_q \tens{I}]^{-1}
= \phi_p\beta_{pq}  \tens{I} - \sum_{n=2}^\infty
\fn{\tens{A}_n^{(p)}}{\vect{k}_q} {\beta_{pq}}^{n},
\label{strong}
\end{align}
where $\fn{\tens{A}_n^{(p)}}{\vect{k}_q}$ is a wave-vector-dependent second-rank tensor that is a functional involving the set of correlation functions $S_1^{(p)},S_2^{(p)},\cdots, S_n^{(p)}$ (defined in Sec. \ref{n-point}) and products of the second-rank tensor field $\fn{\tens{H}^{(q)}}{\vect{r}}$, which is explicitly given as [see also Eq. \eqref{eq:H-tensor-supp}] 
          \begin{align}
          \fn{\ten{H^{(q)}}{ij}}{\vect{r}} =&
          \begin{cases}
          \frac{i}{4}\qty[
                  {k_q}^2 \Hankel{0}{k_q r} - \frac{k_q}{r}
\Hankel{1}{k_q r}]\ten{\delta}{ij}  + \frac{i{k_q}^2}{4} \Hankel{2}{k_q r} \uvect{r}_i
\uvect{r}_j
            ,   & d=2 \\
          \frac{\fn{\exp}{ik_q r}}{\varepsilon_q 4\pi r^3}
                  \big\{
                  \qty[-1+ik_q r +(k_q r)^2]\ten{\delta}{ij} 
                 +\qty[3-3ik_q
r -(k_q r)^2]\uvect{r}_i \uvect{r}_j
                  \big\}
           ,    & d=3
          \end{cases}
          \end{align}
where $k_q \equiv\abs{\vect{k}_q}$. 
Specifically, for $n=2$ and $n\ge 3$, these $n$-point tensors associated with the polarized phase $p$ are, respectively, given by
\begin{align}
\fn{\tens{A}_2^{(p)}}{\vect{k}_q} =& d \varepsilon_q \int_\epsilon \dd{\vect{r}}
\fn{\tens{H}^{(q)}}{\vect{r}}e^{- i \vect{k}_q\cdot \vect{r}}
\fn{\chi_{_V}}{\vect{r}} \label{eq:A2-tensor_direct-space}, \\
\fn{\tens{A}_n^{(p)}}{\vect{k}_q} =&d\varepsilon_q\qty(\frac{-d\varepsilon_q}{\phi_p})^{n-2} \int_\epsilon \dd{\vect{r}_1}
\cdots \dd{\vect{r}_{n-1}} \fn{\tens{H}^{(q)}}{\vect{r}_1- \vect{r}_2}e^{-i
\vect{k}_q\cdot (\vect{r}_1-\vect{r}_2)} \cdot
\fn{\tens{H}^{(q)}}{\vect{r}_2- \vect{r}_3}e^{-i \vect{k}_q\cdot
(\vect{r}_2-\vect{r}_3)} \cdot
          \nonumber\\
          &~\cdots~ \fn{\tens{H}^{(q)}}{\vect{r}_{n-1}- \vect{r}_n}e^{-i
\vect{k}_q\cdot (\vect{r}_{n-1}-\vect{r}_{n})}
\fn{\Delta_n^{(p)}}{\vect{r}_1,\cdots,\vect{r}_n},
\label{eq:An-tensor_direct-space}
\end{align}
where $\int_\epsilon  \dd{\vect{r}}\equiv \lim_{\epsilon\to0^+}\int_{\abs{\vect{r}}>\epsilon}  \dd{\vect{r}}$ and  $\Delta_{n}^{(p)}$ is a position-dependent determinant involving correlation functions of the polarized phase $p$ up to the $n$-point
level:
\begin{equation}\label{eq:determinant}
\fn{\Delta_{n}^{(p)}}{\vect{r}_1,\cdots,\vect{r}_n} =
\begin{vmatrix}
\fn{S_2 ^{(p)}}{\vect{r}_1,\vect{r}_2}      &       \fn{S_1
^{(p)}}{\vect{r}_1} & \cdots & 0 \\
\fn{S_3 ^{(p)}}{\vect{r}_1,\vect{r}_2,\vect{r}_3} & \fn{S_2
^{(p)}}{\vect{r}_2,\vect{r}_3} & \cdots & 0 \\
\vdots & \vdots & \ddots & \vdots \\
\fn{S_n ^{(p)}}{\vect{r}_1,\cdots,\vect{r}_n} &\fn{S_{n-1}
^{(p)}}{\vect{r}_2,\cdots,\vect{r}_n} & \cdots & \fn{S_2
^{(p)}}{\vect{r}_{n-1},\vect{r}_n}
\end{vmatrix}.
\end{equation}

\end{widetext}

For macroscopically isotropic media, the effective dielectric  tensor is isotropic, i.e., $\fn{\tens{\varepsilon}_e}{\vect{k}_q} = \fn{\varepsilon_e}{\vect{k}_q}\tens{I}$.
The corresponding  strong-contrast expansion for $\fn{\varepsilon_e}{\vect{k}_q}$ is obtained by taking the trace of both sides of Eq. \eqref{strong}:
\begin{align}
&{\phi_p}^2 \beta^2_{pq}
[\varepsilon_e(\vect{k}_q) +(d-1)
\varepsilon_q][\varepsilon_e(\vect{k}_q) -
\varepsilon_q]^{-1}
\nonumber\\
=& \phi_p\beta_{pq}   - \sum_{n=2}^\infty
\fn{A_n^{(p)}}{\vect{k}_q} {\beta_{pq}}^{n},
\label{isotropic-strong}
\end{align}
where $\fn{\varepsilon_e}{\vect{k}_q} = \Tr[\fn{\tens{\varepsilon}_e}{\vect{k}_q}]/d$, $\fn{A_n^{(p)}}{\vect{k}_q} = \Tr[\fn{\tens{A}_n^{(p)}}{\vect{k}_q}]/d$ for $n\geq 2$ and  $\Tr[\;]$ denotes the trace operation.
 Furthermore, for statistically isotropic media, the effective dielectric constant becomes independent of the direction of the wave vector, i.e., $\fn{\varepsilon_e}{\vect{k}_q} = \fn{\varepsilon_e}{k_q}$.
\smallskip  

\noindent{\bf Remarks:}
\vspace{-0.1in}

\begin{enumerate}[label=\roman*.]

\item Importantly, the strong-contrast expansion (\ref{strong}) is a series representation of a linear  fractional transformation of
the variable $\tens{\varepsilon}_e({\bf k}_q)$ (left-hand
side of the equation), rather than the effective dielectric constant
tensor itself.
The series expansion in powers of the polarizability $\beta_{pq}$ of this particular rational function of $\tens{\varepsilon}_e({\bf k}_q)$ has important consequences for the predictive power of approximations
derived from the expansion, as detailed
in Sec. \ref{trunc}.

\item  The fact that the exact expansion (\ref{strong}), extracted from our nonlocal relation (\ref{eq:Le_tensor}), is explicitly given in terms of integrals over products of the relevant Green's functions and the $n$-point correlation functions to infinite order implies that multiple-scattering to all orders is exactly treated for the range of wave numbers for which our extended homogenization theory applies, i.e.,  $0 \le |{\bf k}_q| \ell \lesssim 1$. 

\item Note that Eq. \eqref{strong} represents two different series
expansions:~one for $q=1$ and $p=2$ and the other for $q=2$ and $p=1$.

\item The exact expansions represented by (\ref{strong}) are
independent of the reference phase $q$ and hence independent of the wave vector ${\bf k}_q$.

\item For $d=2$,  the strong-contrast expansion applies for TE polarization only.
This implies that the electric field and wave vector are parallel to the plane or transverse to an axis of symmetry in a 3D system whose cross sections are identical.

\item Formally, the original strong-contrast expansions that apply in the quasistatic regime \cite{Re08a} can be obtained from the nonlocal strong-contrast expansions \eqref{strong} by simply replacing the exponential functions  that appear in the expressions for the second-rank tensors
$\fn{\tens{A}_n^{(p)}}{\vect{ k}_q}$, defined by Eq. (\ref{eq:An-tensor_direct-space}), by unity.

\item For statistically isotropic media, the effective phase speed $\fn{c_e}{k_q}$ and attenuation coefficient $\fn{\gamma_e}{k_q}$ are determined by the scalar effective dielectric constant 
$\fn{\varepsilon_e}{k_q}$:
  \begin{align}
  \fn{c_e}{k_q}/c =& {n_e(k_q)}^{-1}=\Re\qty[1/\sqrt{\fn{\varepsilon_e}{k_q}}], \label{speed}\\
  \fn{\gamma_e}{k_q}/c =& {\kappa_e (k_q)}^{-1} = \Im\qty[1/\sqrt{\fn{\varepsilon_e}{k_q}}],
  \label{atten}
  \end{align}
where  ${n_e}(k_q)$ and $\fn{\kappa_e}{k_q}$ are the  effective refractive and extinction indices, respectively.
The quantity $\fn{\exp}{-2\pi \gamma_e /c_e}$ is the factor by which the incident-wave amplitude is attenuated for a period $2\pi/\omega$.

\item The strong-contrast formalism is a significant departure from perturbative expansions obtained from standard
multiple-scattering theory \cite{Fr68,Sh95,Ts01,Ca15}. The operator $\tens{S}$ defined in Eq. (\ref{SO}) is a generalization of the standard scattering operator $\tens{\mathcal{T}}=[\tens{I} - \tens{V}\tens{G}]^{-1}\tens{V}$ \cite{Fr68,Sh95,Ts01},
where $\tens{V}\equiv  \qty[\fn{\varepsilon}{\vect{r}}-\varepsilon_q] \tens{I}$ is the {\it scattering potential}.
Thus, the perturbation series resulting from  $\tens{\mathcal{T}}$ is a weak-contrast expansion with the inherent
limitations that it converges only for small contrast ratios (see also Sec. \ref{trunc}). By contrast, due to the
different possible choices for the exclusion regions and reference phases, 
there is an infinite variety of series expansions that result from the strong-contrast formalism with generally fast-convergence properties. The particular strong-contrast expansion can be designed for different classes of microstructures (see Appendixes \ref{app:exclusion-regions} and D).
The corresponding strong-contrast ``{\it self-energy}" $\tens{\Sigma}$ is a linear fractional transform of $\tens{L}_e$ [see Eq. \eqref{final}],
namely, $\tens{\Sigma}={\tens{L}_e} ^{-1}\qty[\tens{I} -\tens{D} {\tens{L}_e}^{-1}]^{-1}$, which is a
generalization of the self-energy in standard multiple-scattering theory \cite{Fr68,Sh95,Ts01,Ca15}. 
 Thus, it is highly nontrivial to relate diagrammatic expansions of the strong-contrast
formalism to those of multiple-scattering theory.
An elaboration of how strong-contrast expansion generalize those from multiple-scattering theory is presented in the Supplementary Material \cite{SM}.
\label{remark8}
\end{enumerate}

\subsection{Convergence properties and accuracy of truncated series}
\label{trunc}

The form of the strong-contrast expansion parameter $\beta_{pq}$ in Eq. (\ref{strong}) is a direct consequence of the choice of a spherical region excluded from the volume integrals in Eq. (\ref{strong}) due to singularities in the Green's functions \cite{Re08a}.
It is bounded by
\begin{equation} 
-\frac{1}{d-1} \leq \beta_{pq} \equiv \frac{\varepsilon_p
-\varepsilon_q}{\varepsilon_p +  (d-1)\varepsilon_q} \leq 1,
\end{equation}
which implies that the strong-contrast expansion (\ref{strong}) can converge rapidly, even for
infinite contrast ratio $\varepsilon_p/\varepsilon_q \to \infty$.
Other choices for the shape of the exclusion region will lead to different
expansion parameters that will generally be bounded but can lead to 
expansions with significantly different convergence properties \cite{To02a}.
In Appendix \ref{app:exclusion-regions}, we present the corresponding expansions for disk-like and needle-like exclusion regions, which 
are exceptional cases that lead to slowly converging {\it weak-contrast} expansions with expansion parameter  $(\varepsilon_p-\varepsilon_q)/\varepsilon_q$, and thus are unbounded when $\varepsilon_p/\varepsilon_q \to \infty$.

\begin{figure}[bthp]
\centerline{\includegraphics[  width=2.7in,
keepaspectratio,clip=]{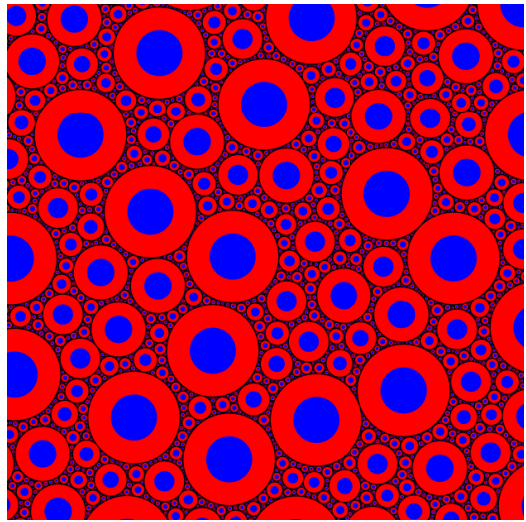}}
\caption{Schematic of the optimal multiscale ``coated-spheres" model that realizes the isotropic Hashin-Shtrikman bounds on $\varepsilon_e$ \cite{Ha63}. 
Each composite sphere is composed of a spherical inclusion of one phase (dispersed phase) that is surrounded by a concentric spherical shell of the other phase such that the fraction of space occupied by the dispersed phase is equal to its overall phase volume fraction. The composite spheres fill all space, implying that their sizes range down to the infinitesimally small. 
When phase 2 is the disconnected inclusion (dispersed) phase, this two-phase medium minimizes and maximizes the effective static dielectric constant $\varepsilon_e$ for prescribed volume fraction and contrast ratio, when $\varepsilon_2/\varepsilon_1 >1$  and $\varepsilon_2/\varepsilon_1< 1$, respectively. 
It has recently been proved that these highly degenerate optimal Hashin-Shtrikman multiscale distributions of spheres are hyperuniform \cite{Ki19a,Ki19b}.
}
\label{HS}
\end{figure}

Importantly, in the purely static case, the expansion (\ref{strong})  becomes identical to one derived by Sen and Torquato \cite{Se89} and its truncation after second-order terms [i.e., setting $\tens{A}_n^{(p)}=0$ for all $n\ge 3$] yields the generalized Hashin-Shtrikman bounds \cite{Ha63} derived by Willis \cite{Wi81a} that are optimal since
they are realized by certain statistically anisotropic composites in which there is a disconnected, dispersed phase in a connected matrix phase \cite{Mi81a}.
In the case of an isotropic effective dielectric constant $\varepsilon_e$, the optimal Hashin-Shtrikman upper and lower bounds for any phase-contrast ratio $\varepsilon_2/\varepsilon_1$ are exactly realized by the multiscale  ``coated-spheres" model, which is depicted in Fig. \ref{HS} in two dimensions. Affine transformations of the coated spheres in the $d$ orthogonal directions lead to oriented coated ellipsoids that are optimal for the
macroscopically anisotropic case.
The lower bound corresponds to the case when the high-dielectric-constant phase is the dispersed, disconnected phase and the upper bound corresponds to the instance in which the high-dielectric-constant phase is the connected matrix. 
Thus, Torquato \cite{To85f,To02a} observed that the strong-contrast expansions (\ref{strong}) in the static limit can be regarded as ones that perturb around such optimal composites, implying that the first few terms of the expansion can yield accurate approximations of the effective property for a class of particulate composites as well as more general microstructures, depending on whether the high-dielectric phase percolates or not.  
For example, even  when  $\varepsilon_2/\varepsilon_1 \gg 1$,  the dispersed phase 2 can consist of identical or polydisperse particles of general shape (ellipsoids, cubes, cylinders, polyhedra) with {\it prescribed orientations}
that may or not {\it overlap}, provided that the particles are prevented from forming large clusters
compared to the specimen size. Moreover, when $\varepsilon_2/\varepsilon_1 \ll 1$, the matrix phase
can be a cellular network \cite{To18c}. Finally, for moderate values of the contrast ratio  $\varepsilon_2/\varepsilon_1$,
even more general microstructures (e.g., those without well-defined inclusions) can be accurately treated. 
Importantly, we show that for the dynamic problem under consideration, the first few terms of the expansion (\ref{strong}) yield accurate approximations of $\tens{\varepsilon}_e({\bf k}_q)$ for a similar wide class of two-phase media (see Sec. \ref{sim-results}). Analogous approximations were derived and applied for the quasistatic regime \cite{Re08a,Ch18a}.

We now show how lower-order truncations of the series (\ref{strong}) can well approximate 
higher-order functionals (i.e., higher-order diagrams) of the exact series to all orders
in terms of lower-order diagrams.
Such truncations of strong-contrast expansions
are tantamount to approximate but resummations of the strong expansions, which enables
multiple-scattering and spatial dispersion effects to be accurately captured to all orders.
Solving the left-hand side of Eq. (\ref{strong}) for $\tens\varepsilon_e$
yields the rational function in $\beta_{pq}$:
\begin{align}
\frac{\fn{\tens{\varepsilon}_e}{\vect{k}_q}}{\varepsilon_q}
   = & \tens{I} + d{\phi_p}^2\beta_{pq} \Big[\phi_p (1-\phi_p 
\beta_{pq})\tens{I} \nonumber \\
&-\sum_{n=2}^\infty \fn{\tens{A}_n^{(p)}}{\vect{k}_q} {\beta_{pq}}^{n-1}\Big]^{-1}.
\label{strong-2}
\end{align}
Expanding Eq. (\ref{strong-2}) in powers of the scalar polarizability
$\beta_{pq}$ yields the series 
\begin{equation}        \label{eq:strong-2-2}
\frac{\fn{\tens{\varepsilon}_e}{\vect{k}_q}}{\varepsilon_q}  =
\sum_{n=0}^{\infty} \fn{\tens{B}_n^{(p)}}{\vect{k}_q} {\beta_{pq}}^n,
\end{equation}
where the first several functionals $\fn{\tens{B}_n^{(p)}}{\vect{k}_q} $ are
explicitly given in terms of $\tens{A}_0^{(p)},~\tens{A}_1^{(p)}, \ldots,~\tens{A}_n^{(p)}$ as
\begin{align*}
\fn{\tens{B}_0^{(p)}}{\vect{k}_q}  =& \tens{I} \\
\fn{\tens{B}_1^{(p)}}{\vect{k}_q}  =& d \phi_p \tens{I}\\
        \fn{\tens{B}_2^{(p)}}{\vect{k}_q}
=& d\qty[\fn{\tens{A}_2^{(p)}}{\vect{k}_q} + \phi_p^2 \tens{I}]\\
        \fn{\tens{B}_3^{(p)}}{\vect{k}_q}
=&
\frac{d}{\phi_p}\Big[{\fn{\tens{A}_2^{(p)}}{\vect{k}_q}}^2 + \phi_p
\fn{\tens{A}_3^{(p)}}{\vect{k}_q}  \nonumber \\ 
&\quad + 2\phi_p^2 \fn{\tens{A}_2^{(p)}}{\vect{k}_q}   +
\phi_p^4 \tens{I}\Big]\\
\fn{\tens{B}_4^{(p)}}{\vect{k}_q}
=&
\frac{d}{\phi_p^2} \Bigg[\fn{\tens{A}_2^{(p)}}{\vect{k}_q}^3+ 2 \phi_p
\fn{\tens{A}_2^{(p)}}{\vect{k}_q} \cdot \fn{\tens{A}_3^{(p)}}{\vect{k}_q} 
    \nonumber \\
    &~+3 \phi_p^2
{\fn{\tens{A}_2^{(p)}}{\vect{k}_q}}^2
   + \phi_p^2 \fn{\tens{A}_4^{(p)}}{\vect{k}_q} 
    \nonumber \\
    &~+ 2 \phi_p^3
\fn{\tens{A}_3^{(p)}}{\vect{k}_q} +3  \phi_p^4 \fn{\tens{A}_2^{(p)}}{\vect{k}_q}  +
\phi_p^6 \tens{I}\Bigg],
\end{align*}
where $\tens{T}^n$ stands for $n$ successive inner products of a second-rank tensor $\tens{T}$.

Let us now compare the exact expansion \eqref{eq:strong-2-2} to the one
that results when expanding the truncation of the exact expression (\ref{strong}) for $[\tens{\varepsilon}_e +(d-1) \tens{\varepsilon}_q]\cdot(\tens{\varepsilon}_e -\tens{\varepsilon}_q)^{-1}$ at the two-point level:
\begin{align}\label{eq:two-point}
\frac{\fn{\tens{\varepsilon}_e}{\vect{k}_q}}{\varepsilon_q}
   \approx & \tens{I} +
d{\phi_p}\beta_{pq}\qty[{(1-\phi_p\beta_{pq})\tens{I} -
\fn{\tens{A}_2^{(p)}}{\vect{k}_q} \beta_{pq}/\phi_p}]^{-1}\\
   = & \sum_{n=0}^{\infty} \fn{\tens{C}^{(p)}_n}{\vect{k}_q} {\beta_{pq}}^n,
\label{strong-3}
\end{align}
where the $n$th-order functional $\fn{\tens{C}^{(p)}_n}{\vect{k}_q}$ for any $n$ is given 
in terms of the volume fraction $\phi_p$ and $\fn{\tens{A}_2^{(p)}}{\vect{k}_q}$, which has the following diagrammatic representation:
\begin{equation}
\fn{\tens{A}_2^{(p)}}{\vect{k}_q}= 
\begin{minipage}[h]{1.0in}
\includegraphics[width=1.in]{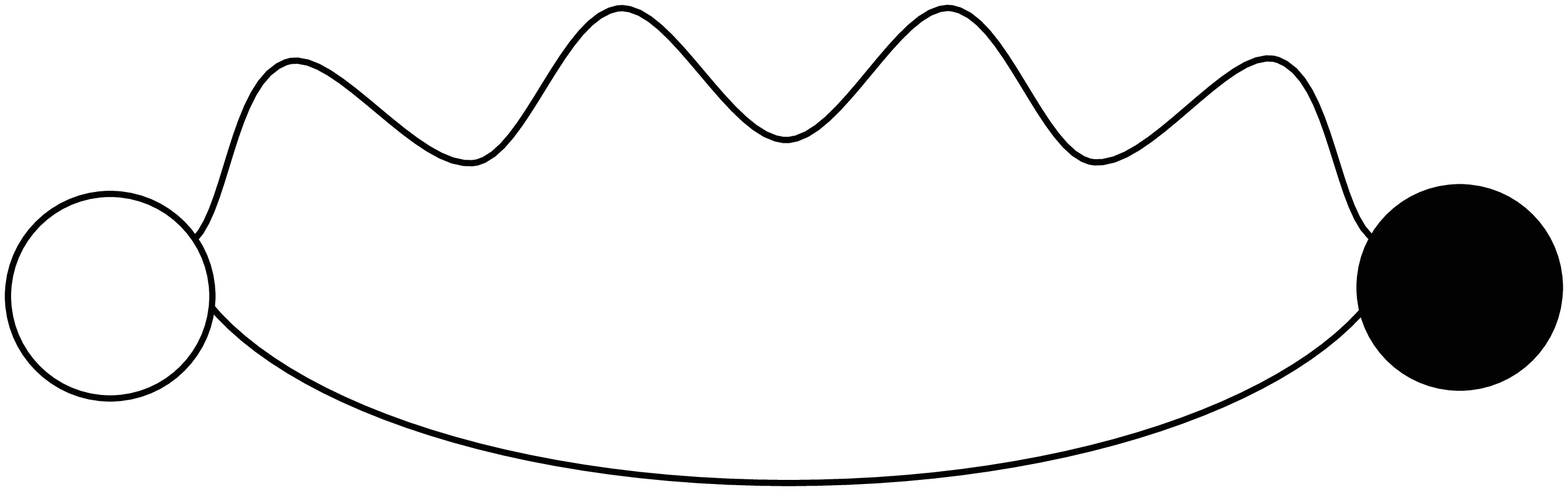}
\end{minipage}.
\label{A2}
\end{equation}
\noindent{Here the solid and wavy lines joining two nodes represent the spatial
correlation via  $\fn{\chi_{_V}}{\vect{r}}$ and a wave vector-dependent Green's function $\fn{\tens{H}^{(q)}}{\vect{r}}e^{- i \vect{k}_q\cdot \vect{r}}$ between the nodes,
respectively. The black node indicates a volume integral and carries a factor of  $d\varepsilon_q$.
The first several functionals $\fn{\tens{C}^{(p)}_n}{\vect{k}_q}$ are
explicitly given as}
\begin{align*}
\fn{\tens{C}^{(p)}_0}{\vect{k}_q} & = \tens{I}, \\
\fn{\tens{C}^{(p)}_1}{\vect{k}_q} & = d \phi_p \tens{I},\\
\fn{\tens{C}^{(p)}_2}{\vect{k}_q} & = d \qty[\fn{\tens{A}_2^{(p)}}{\vect{k}_q} +
{\phi_p}^2\tens{I}],\\
\fn{\tens{C}^{(p)}_3}{\vect{k}_q} & = \frac{d}{\phi_p}
\qty[\fn{\tens{A}_2^{(p)}}{\vect{k}_q} + {\phi_p}^2\tens{I}]^2,\\
\fn{\tens{C}^{(p)}_4}{\vect{k}_q} & = \frac{d}{{\phi_p}^2}
\qty[\fn{\tens{A}_2^{(p)}}{\vect{k}_q} + {\phi_p}^2\tens{I}]^3.
\end{align*}
Thus, comparing (\ref{eq:strong-2-2}) to (\ref{strong-3}), we see that
truncation of the expansion (\ref{strong}) for $[\tens{\varepsilon}_e
+(d-1) \tens{\varepsilon}_q]\cdot(\tens{\varepsilon}_e -
\tens{\varepsilon}_q)^{-1}$
at the two-point level actually translates into approximations of the
higher-order functionals to all orders in terms of the first-order diagram $\phi_p$ and the second-order
diagram (\ref{A2}).
This two-point truncation can be thought of as an approximate but accurate resummed representation of the exact expansion (\ref{eq:strong-2-2}).
Note that the approximate expansion (\ref{strong-3}) is exact through second order in
$\beta_{pq}$.
Clearly, truncation of Eq. (\ref{strong}) at the three-point level (see Appendix \ref{3-pt})
will yield even better approximations of the higher-order functionals.

\section{Strong-Contrast Approximation Formulas} \label{sec:strong-contrast}

Here we describe lower-order truncations of the strong-contrast expansions that are expected to yield accurate closed-form formulas for $\tens{\varepsilon}_e(\vect{k}_q)$ that apply
over a broad range of  wavelengths   ($k_q\ell \lesssim 1$), volume fractions and contrast ratios for a wide class of microstructures.

\subsection{Macroscopically anisotropic media}

For the ensuing treatment, it is convenient to rewrite the expansions \eqref{strong}, valid for macroscopically anisotropic media in $\R^d$, in the following manner:
\begin{align}	&{\phi_p}^2 \beta^2_{pq}
\frac{\tens{\varepsilon}_e(\vect{k}_q) +(d-1)
\varepsilon_q\tens{I}}{\tens{\varepsilon}_e(\vect{k}_q) -
\varepsilon_q \tens{I}} 
\nonumber \\
=& \phi_p\beta_{pq}  \tens{I} - \sum_{n=2}^M
\fn{\tens{A}_n^{(p)}}{\vect{k}_q} {\beta_{pq}}^n + \fn{\mathcal{R}_M}{\vect{k}_q}, \label{eq:expansions_remainder}
\end{align}
where the $M$th-order remainder term is defined as
\begin{equation} \label{eq:remainder}
\fn{\mathcal{R}_M}{\vect{k}_q} \equiv \sum_{n=M+1}^\infty
\fn{\tens{A}_n^{(p)}}{\vect{k}_q} {\beta_{pq}}^n.
\end{equation}
Truncating the exact nonlocal expansion \eqref{eq:expansions_remainder} at the two- and three-point levels, i.e., setting $\fn{\mathcal{R}_2}{\vect{k}_q}=0$ and $\fn{\mathcal{R}_3}{\vect{k}_q}=0$, respectively, yields 
\begin{align}	\label{eq:two-point-approx_anisotropic}
{\phi_p}^2 \beta^2_{pq}
\frac{\tens{\varepsilon}_e(\vect{k}_q) +(d-1)
\varepsilon_q\tens{I}}{\tens{\varepsilon}_e(\vect{k}_q) -
\varepsilon_q \tens{I}}
=& \phi_p\beta_{pq}  \tens{I} - 
\fn{\tens{A}_2^{(p)}}{\vect{k}_q} {\beta_{pq}}^2, 
\\
{\phi_p}^2 \beta^2_{pq}
\frac{\tens{\varepsilon}_e(\vect{k}_q) +(d-1)
\varepsilon_q\tens{I}}{\tens{\varepsilon}_e(\vect{k}_q) -
\varepsilon_q \tens{I}}
=& \phi_p\beta_{pq}  \tens{I} - \Big[\fn{\tens{A}_2^{(p)}}{\vect{k}_q} {\beta_{pq}}^{2} 
\nonumber \\
&\quad+ \fn{\tens{A}_3^{(p)}}{\vect{k}_q} {\beta_{pq}}^{3}\Big], \label{eq:three-point-approx_anisotropic}
\end{align}
Compared to the quasistatic approximation \cite{Re08a}, these nonlocal approximations substantially extend the range of applicable wave number, namely, $0\leq \abs{\vect{k}_q} \ell \lesssim 1$.

\subsection{Macroscopically isotropic media}
\label{strong-approx}

All of the applications considered in this paper, will focus on the case of macroscopically isotropic media, i.e., they are described by the scalar effective dielectric constant $\fn{\varepsilon_e}{\vect{k}_q}= \mbox{Tr}\, [\tens{\varepsilon}_e(\vect{k}_q)]/d$ but depend on the direction of the wave vector $\vect{k}_q$.

\subsubsection{Strong-contrast approximation at the two-point level}

Solving Eq. \eqref{isotropic-strong} for the  effective dielectric constant $\fn{\varepsilon_e}{\vect{k}_q}$ yields the strong-contrast approximation for macroscopically isotropic media:
\begin{align}
\frac{\fn{\varepsilon_e}{\vect{k}_q}}{\varepsilon_q} &= 1+ \frac{d{\beta_{pq}} {\phi_p} ^2}{\phi_p(1-\beta_{pq} \phi_p) -\beta_{pq} \fn{A_2^{(p)}}{\vect{k}_q}} 
\nonumber \\
&= 
1+ \frac{d{\beta_{pq}} {\phi_p} ^2}{\phi_p(1-\beta_{pq} \phi_p) + \frac{(d-1)\pi \beta_{pq}}{2^{d/2}\fn{\Gamma}{d/2}} \fn{F}{\vect{k}_q} },
\label{eq:modified-strong-contrast-approximation_2pt}
\end{align}
where $\beta_{pq}$ is defined in Eq. \eqref{eq:polarizability}, $\fn{A_2 ^{(p)}}{\vect{k}_q} \equiv \Tr[\fn{\tens{A}_2^{(p)}}{\vect{k}_q}]/d$, and  $\fn{F}{\vect{Q}}$ is what we call the {\it nonlocal attenuation function} of a composite for reasons we describe below.
The direct- and Fourier-space representations of $\fn{F}{\vect{Q}}$ are given as
\begin{align}
\fn{F}{\vect{Q}} &\equiv - \frac{2^{d/2} \fn{\Gamma}{d/2}}{\pi}{Q}^2 \int_\epsilon \frac{i}{4}\qty(\frac{Q}{2\pi r})^{d/2-1}
\nonumber \\
&\quad\quad\times \Hankel{d/2-1}{Q r}e^{-i\vect{Q} \cdot \vect{r}}\fn{\chi_{_V}}{\vect{r}} \dd{\vect{r}} \label{eq:attenuation-function} 
\\
&=
-\frac{ \fn{\Gamma}{d/2}}{2^{d/2}\pi^{d+1}}{Q}^2 \int  \frac{\spD{\vect{q}}}{\abs{\vect{q}+\vect{Q}}^2 - Q^2}
\dd{\vect{q}}.\label{eq:attenuation-function_Fourier}
\end{align}
The exponential  $\fn{\exp}{-i\vect{Q} \cdot \vect{r}}$  in Eq. \eqref{eq:attenuation-function} arises from the phase difference associated with the incident waves at positions separated by $\vect{r}$. In the quasistatic regime, this phase factor is negligible, and Eq. \eqref{eq:attenuation-function} reduces to the local attenuation function $\fn{\mathcal{F}}{Q}$ (derived in Ref. \cite{Re08a}  and summarized in the Supplementary Material \cite{SM}) because it is barely different from unity over the correlation length associated with the autocovariance function $\fn{\chi_{_V}}{r}$.
The strong-contrast approximation (\ref{eq:modified-strong-contrast-approximation_2pt}) was postulated in Ref. \cite{Ki20a} on physical grounds. By contrast, the present work derives it as a consequence of our exact nonlocal formalism (Sec. \ref{exact}).

For statistically isotropic media, the effective dielectric constant as well as the attenuation function are independent of the direction of the incident wave vector $\vect{k}_q$, and thus, they can be considered as functions of the wave number, i.e., $\fn{\varepsilon_e}{k_q}=\fn{\varepsilon_e}{\vect{k}_q}$ and $\fn{F}{k_q} = \fn{F}{\vect{k}_q}$.
Then, the real and imaginary parts of Eq. \eqref{eq:attenuation-function} can be simplified as
\begin{align}
\Im[\fn{F}{Q}] =&
\begin{cases}
- \frac{Q^2}{\pi^2} \int_0^{\pi/2} \spD{2Q \cos\phi}\dd{\phi}, & d=2 \\
- \frac{Q}{2(2\pi)^{3/2}} \int_0^{2Q} q\spD{q}\dd{q}, & d=3  
\end{cases} \label{eq:ImF_dD} 
\\
\Re[\fn{F}{Q}] =& 
-\frac{2 Q^2}{\pi } 
\mathrm{p.v.}  \int_{0}^{\infty} \dd{q}\frac{1}{q(Q^2-q^2)}\Im[\fn{F}{q}],\label{eq:ReF_dD}
\end{align}
where Eq. \eqref{eq:ReF_dD} is valid for $d=2,3$, and $\mathrm{p.v.}$ stands for the Cauchy principal value.
Following conventional usage, we say that a composite attenuates waves at a given wave number if the imaginary part of the effective dielectric constant is positive. Recall that attenuation in the present study occurs only because of multiple-scattering effects (not absorption).
While it is the imaginary part of $F(Q)$ that determines directly the degree of attenuation or, equivalently, $\Im[\varepsilon_e]$, we see from Eq. (\ref{eq:ReF_dD}) that the real part of $F (Q)$ is directly related to its imaginary part.
It is for this reason that we refer to the complex function $F(Q)$ as the (nonlocal) attenuation function.

\subsubsection{Modified  strong-contrast approximation at the two-point level}

Here we extend the validity of the strong-contrast approximation \eqref{eq:modified-strong-contrast-approximation_2pt} so that it is accurate at larger wave numbers and hence better captures spatial dispersion. 
This is done by an appropriate rescaling of the wave number in the reference phase, $k_q$, which we show is tantamount to approximately accounting for higher-order contributions in the remainder term $\fn{\mathcal{R}_2}{{k}_q}$.
Given that the strong-contrast expansion for  isotropic media perturbs around the Hashin-Shtrikman structures (see Fig. \ref{HS}) in the static limit, it is natural to use the scaling  $\sqrt{\varepsilon_\mathrm{HS} /\varepsilon_q}k_q$, where   $\varepsilon_\mathrm{HS}$ is the Hashin-Shtrikman estimate, i.e.,
\begin{equation} \label{eq:HS-bound}
\varepsilon_\mathrm{HS} \equiv \varepsilon_q \qty[1+\frac{d\phi_p\beta_{pq}}{1-\phi_p \beta_{pq}}],  
\end{equation}
which gives the Hashin-Shtrikman lower bound and upper bound if $\varepsilon_p > \varepsilon_q$ and  $\varepsilon_p < \varepsilon_q$, respectively.
This scaling yields the following {\it scaled} strong-contrast approximation for statistically isotropic media:
\begin{equation} \label{eq:scaled-2pt-approx}
\frac{\fn{\varepsilon_e}{k_q}}{\varepsilon_q} = 1 +  \frac{d{\beta_{pq}} {\phi_p} ^2}{\phi_p(1-\beta_{pq} \phi_p) + \frac{(d-1)\pi \beta_{pq}}{2^{d/2}\fn{\Gamma}{d/2}} \fn{F}{\sqrt{\frac{\varepsilon_\mathrm{HS}}{\varepsilon_q}}k_q }}.
\end{equation}

We now show that the scaled approximation \eqref{eq:scaled-2pt-approx} indeed provides good estimates of leading-order corrections of $\fn{\mathcal{R}_2}{{k}_q}$ in powers of $k_q$.
To do so, we employ the concept of the averaged (effective) Green's function $\E{\fn{\tens{G}^{(q)}}{\vect{q}}}$ of an inhomogeneous medium which in principle accounts for the all multiple-scattering events
\begin{align}
\E{\fn{\tens{G}^{(q)}}{\vect{q}}} = \qty(\frac{\omega}{c})^2 \qty{\qty[q^2-{\fn{k_e}{\omega}}^2]\tens{I} -\vect{q}\vect{q}}^{-1}, \label{eq:eff.GreenFn}
\end{align}
where ${\fn{k_e}{\omega}} \equiv \sqrt{\fn{\varepsilon_e}{\omega}} \omega/c = \sqrt{\fn{\varepsilon_e}{\omega}/\varepsilon_q}k_q$ is the effective wave number at a frequency $\omega$,
and $\fn{\varepsilon_e}{\omega}$ is the exact effective dynamic dielectric constant, assuming a well-defined homogenization description \cite{Sh95, Ca09b}.
Since exact complete microstructural information is, in principle, accounted for with the  effective Green's function (\ref{eq:eff.GreenFn}), the exact strong-contrast expansion can be approximately equated to  the one  truncated at the two-point level with  an attenuation function given in terms of the effective Green's function, i.e.,
\begin{align}	
&{\phi_p}^2 \beta^2_{pq}
\frac{\varepsilon_e(k_q) +(d-1)
\varepsilon_q}{\varepsilon_e(k_q) -
\varepsilon_q}
	\nonumber \\
	=& 
\phi_p\beta_{pq} - \fn{A_2^{(p)}}{k_q} {\beta_{pq}}^2 +  \fn{\mathcal{R}_2}{k_q}, \label{eq:effective-expansions}
	\\
	\approx&
\phi_p\beta_{pq} - \fn{A_2^{(p)}}{\fn{k_e}{\omega}} {\beta_{pq}}^2. \label{eq:expansions-by-averaged-GreenFn}
\end{align}
When the functional form of $\fn{A_2^{(p)}}{Q}$ or, equivalently, $\fn{F}{Q}$ is available, it is possible to solve Eq. \eqref{eq:expansions-by-averaged-GreenFn} for $\fn{\varepsilon_e}{k_q}$ in a self-consistent manner. Instead, we show that by assuming $k_e(\omega) \approx \sqrt{\varepsilon_\mathrm{HS}/\varepsilon_q}k_q$, which results in Eq. \eqref{eq:scaled-2pt-approx},   we obtain  good estimates of the leading-order corrections of $\fn{\mathcal{R}_2}{{k}_q}$ in powers of $k_q$. 
Thus,  the scaled approximation \eqref{eq:scaled-2pt-approx} provides better estimates of the higher-order three-point approximation (given in Appendix \ref{3-pt}) than the unmodified strong-contrast approximation \eqref{eq:two-point}.
This can be easily confirmed in the quasistatic regime from the small-$k_q$ expansions of $\fn{A_2^{(p)}}{k_q}$ and $\fn{A_3^{(p)}}{k_q}$ given in Ref. \cite{Re08a}.
We also confirm the improved predictive capacity of the scaled strong-contrast approximation to better capture dispersive characteristics for ordered and disordered models via comparison to finite-difference time-domain (FDTD) simulations; see Figs. \ref{fig:periodic-simulations} and \ref{fig:3D-disordered-simulations}, and Sec. VIII in the Supplementary Material \cite{SM}.

\section{Model Microstructures}\label{sec:models}

\begin{figure*}[t]
\includegraphics[width = 0.92\textwidth]{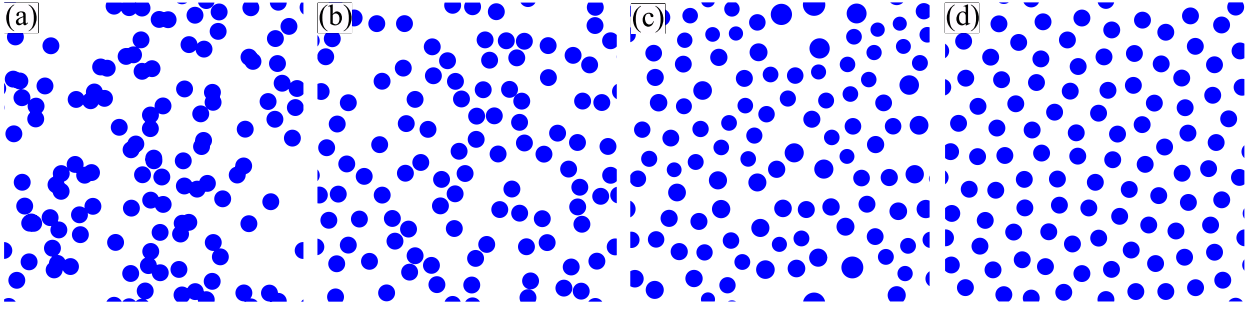}
\caption{Representative images of configurations of the four models of 2D disordered particulate media described in this section. These include
(a) overlapping spheres, (b) equilibrium packings, (c) class I hyperuniform polydisperse packings, and (d) stealthy hyperuniform packings.
For all models, the volume fraction of the dispersed phase (shown in black) is $\phi_2 = 0.25$.
Note that (a) and (b) are not hyperuniform.
While these models consist of distributions of particles, both overlapping and nonoverlapping, the formulas derived in Sec. \ref{sec:strong-contrast} can be applied to any two-phase microstructure. Indeed, Appendix D describes applications to media with phase-inversion symmetry.
\label{fig:model_dispersions}
}
\end{figure*}

We consider four models of 2D and 3D disordered media 
to understand the effect of microstructure on the effective dynamic dielectric constant,
two of which are nonhyperuniform (overlapping spheres and equilibrium hard-sphere packings) and two of which are hyperuniform (hyperuniform polydisperse packings and stealthy hyperuniform packings of identical spheres).
The particles of dielectric constant $\varepsilon_2$ are distributed throughout a matrix of dielectric constant $\varepsilon_1$.
We also compute the spectral density for each model, which is the required microstructural information to evaluate the nonlocal strong-contrast approximations discussed in Sec. VB.

Representative images of configurations of the four aforementioned models of 2D disordered particulate media are depicted in Fig. \ref{fig:model_dispersions}.
Note that the degree of volume-fraction fluctuations decreases from the leftmost image to the rightmost one.

\subsection{Overlapping spheres}\label{sec:overlapping-spheres}

Overlapping spheres (also called the fully-penetrable-sphere model) refer to 
an uncorrelated (Poisson) distribution of spheres of radius $a$ throughout a matrix \cite{To02a}.
For such nonhyperuniform models at number density $\rho$ in $d$-dimensional Euclidean space $\mathbb{R}^d$, the autocovariance function is known analytically \cite{To02a}:     
\begin{equation}\label{eq:autoco_OS}
\fn{\chi_{_V}}{r} = 
\exp\qty(-\rho   v_2(r;a))-{\phi_1}^2,
\end{equation}
where $\phi_1 = \exp(-\rho\fn{v_1}{a} )$ is the volume fraction of the matrix phase (phase 1), $v_1(a)$ is given by (\ref{v1}), 
and  $\fn{v_2}{r;a}$ represents the union volume of two spheres whose centers are separated by a distance  $r$. 
In two and three dimensions, the latter quantity is explicitly given, respectively, by
\begin{align*}
\frac{\fn{v_2}{r;a}}{\fn{v_1}{a}} = 
\begin{cases} 
    \begin{aligned}
 2 \fn{\Theta}{x-1}
&+ \frac{2}{\pi} \Big[\pi + x \qty(1-x^2)^{1/2} \\
&-\fn{\cos^{-1}}{x} \Big] \fn{\Theta}{1-x}  
    \end{aligned}
, & d=2 \\
2 \fn{\Theta}{x-1}+\qty(1+\frac{3x}{2}-\frac{x^3}{2}) \fn{\Theta}{1-x} , & d=3
\end{cases}
\end{align*}
where $x\equiv r/2a$, and $\Theta(x)$ (equal to 1 for $x>0$ and zero otherwise) is the Heaviside step function.
For $d=2$ and $d=3$, the particle phase (phase 2) percolates when $\phi_2 \approx 0.68$ (Ref. \cite{Qu00}) and $\phi_2 \approx 0.29$
(Ref. \cite{Ri97b}), respectively. The corresponding spectral densities are easily found numerically by performing the Fourier transforms
indicated in Eq. (\ref{def-spec}). In this work, we apply this model 
for $\phi_2$'s well below the percolation thresholds.

\subsection{Equilibrium packings}

Another disordered nonhyperuniform model we treat is distributions of equilibrium (Gibbs) of identical hard spheres of radius $a$ 
along the stable fluid branch \cite{Han13, To02a}. The structure factors of such packings are well approximated by the Percus-Yevick solution \cite{Han13, To02a},
which is analytically solvable for odd values of $d$.   For $d=3$, the Percus-Yevick solution gives the following expression for the structure factor $\fn{S}{Q}$ \cite{To02a}:
\begin{align}
\fn{S}{Q} = 
&\Big(1-\rho \frac{16 \pi  a^3 }{q^6} 
\Big\{\big[24 a_1 \phi_2 - 12 (a_1 + 2 a_2) \phi_2 q^2 
    \nonumber \\
    &\quad+ (12 a_2 \phi_2 + 2 a_1 + a_2\phi_2) q^4] \cos(q) 
    \nonumber \\
&+ [24 a_1 \phi_2 q - 2 (a_1 + 2 a_1 \phi_2 + 12 a_2 \phi_2) q^3\big] \sin(q)
    \nonumber \\
& -24 \phi_2 (a_1 - a_2 q^2) \Big\}\Big)^{-1},
\end{align}
where $q = 2Qa$, $a_1 = (1+2\phi_2)^2/(1-\phi_2)^4$, and $a_2 = -(1+0.5\phi_2)^2 /(1-\phi_2)^4$.
Using this solution in conjunction with Eq. (\ref{chi-packing}) yields the corresponding spectral density ${\tilde \chi}_{_V}(Q)$.
For $d=2$, we obtain the spectral density from Monte Carlo generated disk packings \cite{To02a}.






\subsection{Hyperuniform polydisperse packings}

Class I hyperuniform packings of spheres with a polydispersity in size can be constructed from nonhyperuniform progenitor point patterns via a tessellation-based procedure \cite{Ki19a,Ki19b}.
Specifically, we employ the centers of 2D and 3D configurations of identical hard spheres in equilibrium at a packing  fraction $0.45$ and particle number $N=1000$
as the progenitor point patterns.
One begins with the Voronoi tessellation \cite{To02a}  of these progenitor point patterns.
We then rescale the particle in the $j$th Voronoi cell $C_j$ without changing its center such that the packing fraction inside this cell is identical to a prescribed value $\phi_2<1$.
The same process is repeated over all cells.
The final packing fraction is $\phi_2 = \sum_{j=1}^N \fn{v_1}{a_j}/V_\mathfrak{F} = \rho \fn{v_1}{a}$, where $\rho$ is the number density of particle centers
and $a$ represents the mean sphere radius.
In the small-$\abs{\vect{Q}}$ regime, the spectral densities of the resulting particulate composites exhibit a power-law scaling $\spD{\vect{Q}} \sim \abs{\vect{Q}}^4$,
which are of class I.

\subsection{Stealthy hyperuniform packings}
\label{stealth}

Stealthy hyperuniform particle systems, which are also class I, are defined by the spectral density vanishing around the origin, i.e., $\spD{\vect{Q}}=0$ for $0<\abs{\vect{Q}}\leq Q_\mathrm{U}$;
see Eq.(\ref{stealthy}).    
We obtain the spectral density from realizations of disordered stealthy hyperuniform packings for $d=2,3$ that are numerically generated via the following two-step procedure.
Specifically, we first generate such point configurations consisting of $N$ particles in a fundamental cell $\mathfrak{F}$ under periodic boundary conditions via the collective-coordinate optimization technique \cite{Uc04b,Ba08,Zh15a}, which amounts to finding numerically the ground-state configurations for the following potential energy;
\begin{equation}\label{eq:CC_potential}
\fn{\Phi}{\vect{r}^N} =\frac{1}{V_\mathfrak{F}} \sum_{\vect{Q}} \fn{\tilde{v}}{\vect{Q}}\fn{S}{\vect{Q}} +  \sum_{i <j} \fn{u}{r_{ij}},
\end{equation}
where   
\begin{equation}
\fn{\tilde{v}}{\vect{Q}}
=
\begin{cases}
1, & Q_\text{L}< \abs{\vect{Q}} \leq Q_\text{U}, \\
0, &\mathrm{otherwise},
\end{cases} \label{eq:cc-}
\end{equation}
and a soft-core repulsive term \cite{Zh17a} 
\begin{equation} \label{eq:soft}
\fn{u}{r}
=
\begin{cases}
(1-r/\sigma)^2, & r < \sigma,\\
0,&\mathrm{otherwise}.
\end{cases}
\end{equation}
In contrast to the usual collective-coordinate procedure \cite{Uc04b,Ba08,Zh15a}, the interaction \eqref{eq:CC_potential} used here also includes a soft-core repulsive energy \eqref{eq:soft}, as done in Ref. \cite{Zh17a}.
    Thus, the associated ground-state configurations are still disordered, stealthy and hyperuniform, and their nearest-neighbor distances are larger than the length scale $\sigma$ due to the soft-core repulsion $\fn{u}{r}$.
Finally, to create packings, we follow Ref. \cite{Zh16b} by circumscribing the points by identical spheres of radius $a<\sigma/2$ under the constraint that they cannot overlap
(See  the Supplementary Material \cite{SM} for certain results concerning stealthy ``nonhyperuniform" packings in which $Q_\text{L} >0$.)
The parameters used to generate these disordered stealthy packings are summarized in the Supplementary Material \cite{SM}.

\subsection{Spectral Densities for the Four Models}\label{sec:microstr-chiv}

\begin{figure}[h]
\includegraphics[width = 0.5\textwidth]{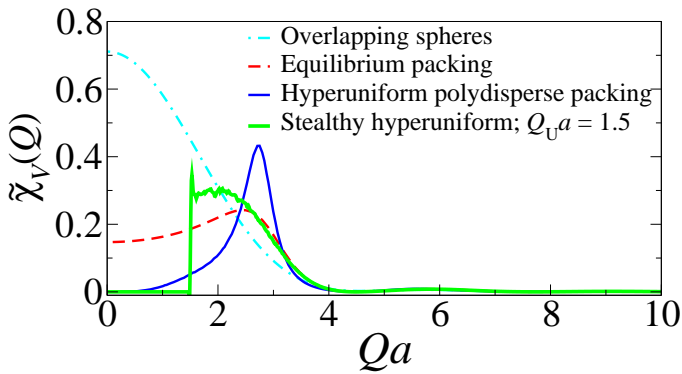}
\caption{The spectral density $\spD{Q}$ as a function of  dimensionless wave number $Qa$ for the four models of
3D disordered media: overlapping spheres, equilibrium packings, class I hyperuniform polydisperse packings, and stealthy hyperuniform packings.
In all cases, the volume fraction of the dispersed phase is $\phi_2 = 0.25$.
For hyperuniform polydisperse packings, $a$ is the mean sphere radius.  The other three models consist of identical spheres of radius $a$.
Corresponding graphs of the spectral densities for the  2D models are provided in the Supplementary Material \cite{SM}.
\label{fig:spectral-density_3D-models}}
\end{figure}

Here, we plot the spectral density $\spD{Q}$ for the four  models  at the selected particle-phase volume fraction of $\phi_2=0.25$; see Fig. \ref{fig:spectral-density_3D-models}.
From the long- to intermediate-wavelength regimes ($Qa \lesssim 4$), their spectral densities are considerably different from one another.
Overlapping spheres depart the most from hyperuniformity, followed by equilibrium packings. 
Stealthy packings suppress volume-fraction fluctuations to a greater degree than hyperuniform polydisperse
packings over a wider range of wavelengths. 
In the small-wavelength regime ($Qa \gg 4$), all four curves tend to collapse onto a single curve, reflecting the fact that all four models
are composed of spheres of similar sizes.

\section{Results for the Nonlocal Attenuation Function} 
\label{attenuation}

We report some  general behaviors of the nonlocal attenuation function $\fn{F}{\vect{Q}}$ [cf. Eqs. (\ref{eq:attenuation-function}) or (\ref{eq:attenuation-function_Fourier})]
for nonhyperuniform and hyperuniform media for long and intermediate wavelengths.
We also provide plots of both the real and imaginary parts of  $\fn{F}{Q}$ 
for the four models of disordered two-phase media considered in this work, which depends on wave number $Q$.

The function $\fn{F}{Q}$ depends on the microstructure via the spectral density $\spD{Q}$.
Thus, assuming that the latter quantity has the power-law scaling $\spD{Q}\sim Q^\alpha$ as $Q\to 0$,
the asymptotic behavior of $\fn{F}{Q}$  in the long-wavelength limit ($Q\to 0$) follows as
\begin{align}
\Im[\fn{F}{Q}] &\sim 
\begin{cases}
Q^d,            &\text{nonhyperuniform}\\
Q^{d+\alpha},   &\text{hyperuniform}
\end{cases},&\text{as~}Q\to 0, \label{eq:small-Q-Im_atten_fn}
\\
\Re[\fn{F}{Q}] &\sim Q^2,& ~~\text{as~}Q\to 0,
\label{eq:small-Q-Re_atten_fn}
\end{align}
where we use Eqs.  (\ref{eq:attenuation-function}) and \eqref{eq:ImF_dD}. Recall that the exponent $\alpha$ lies in the open interval $(0,\infty)$ for hyperuniform systems (see Sec. \ref{hyper}). For nonhyperuniform systems studied here, we take $\alpha=0$.
The reader is referred to the Supplementary Material \cite{SM} for derivations of Eqs. (\ref{eq:small-Q-Im_atten_fn}) and  \eqref{eq:small-Q-Re_atten_fn}.
Importantly, in the quasistatic regime, the imaginary parts of the effective dielectric constant for 
both strong-contrast approximations, \eqref{eq:modified-strong-contrast-approximation_2pt} and 
\eqref{eq:scaled-2pt-approx}, are determined by the asymptotic behaviors $\fn{F}{Q}$ indicated in Eq. (\ref{eq:small-Q-Im_atten_fn}), i.e.,
\begin{align}
&\Im[\fn{\varepsilon_e}{k_q}] \sim \Im[\fn{F}{k_q}]
\nonumber \\
\sim&
\begin{cases}
{k_q}^d,                    &\text{nonhyperuniform}\\
{k_q}^{d+\alpha},   &\text{hyperuniform}
\end{cases},\quad\text{as~}k_q\to 0. \label{eq:Im-diel-quasistatic}
\end{align}
Thus, hyperuniform media are less lossy than their nonhyperuniform counterparts in the quasistatic regime.


In the case of stealthy  hyperuniform media [i.e., $\spD{Q}=0$ for $0 \le Q < Q_\mathrm{U}$], 
the imaginary part of $\fn{F}{Q}$ defined by Eq. \eqref{eq:ImF_dD} is identically zero  (transparent or lossless) for any
space dimension $d$ for a range of wave numbers; specifically,
\begin{align}
\Im[\fn{F}{Q}] = 0, ~~\text{for } 0 \le Q< Q_\mathrm{U}/2.
\label{eq:transparency-regime}
\end{align}
(The Supplementary Material \cite{SM} describes how the  local attenuation function $\fn{\mathcal{F}}{Q}$ derived in Ref. \cite{Re08a} 
generally differs from its nonlocal counterpart.) 
The transparency interval in which $\Im[\fn{\varepsilon_e}{k_q}]=0$ predicted by the two strong-contrast approximations [Eqs. \eqref{eq:modified-strong-contrast-approximation_2pt} and  \eqref{eq:scaled-2pt-approx}] is thus given by
\begin{align}
&\Im[\fn{\varepsilon_e}{k_q}]=0,\nonumber \\
&\quad\text{for~}
\begin{cases}
0 \le k_q  < \frac{Q_\mathrm{U}}{2},& \text{[from Eq. \eqref{eq:modified-strong-contrast-approximation_2pt}],}  \\
0\le k_q  < \frac{Q_\mathrm{U} }{2 (\varepsilon_\mathrm{HS}/\varepsilon_q)^{1/2}}, &\text{[from Eq. \eqref{eq:scaled-2pt-approx}}],
\end{cases} \label{eq:transparency-regime-strong-contrast}
\end{align}
where $\varepsilon_\mathrm{HS}$ is given in Eq. \eqref{eq:HS-bound}.
When $\varepsilon_p > \varepsilon_q$, since $\varepsilon_\mathrm{HS} >\varepsilon_q$, the scaled approximation 
accurately predicts a narrower transparency interval than the unscaled variant, as verified in Sec. \ref{sim-results}.
Interestingly, the transparency interval obtained from the less 
accurate formula \eqref{eq:modified-strong-contrast-approximation_2pt} agrees  with the one obtained from previous
simulation results for stealthy hyperuniform ``point" scatterers \cite{Le16}, not the finite-sized particles considered here.

\begin{figure}[h]
\includegraphics[width = 0.42\textwidth]{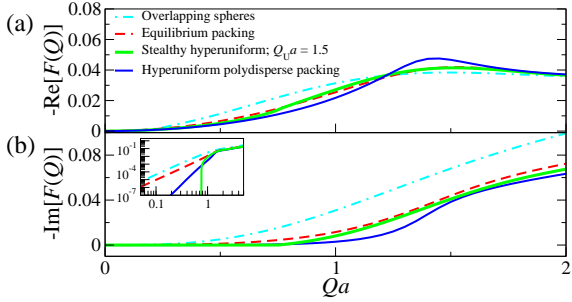}
\caption{The negative values of (a) the real and (b) imaginary parts of the nonlocal attenuation function $\fn{F}{Q}$ as a function of the dimensionless wave number $Qa$ [defined in Eq. \eqref{eq:attenuation-function}] for the four models of 3D disordered composite media considered in this paper.
The inset in (b) is the log-log plot of the larger panel.
The volume fraction of the dispersed phase for each model is $\phi_2 = 0.25$.
The first three models consist of identical spheres of radius $a$.
For class I hyperuniform polydisperse particulate media, $a$ is the mean sphere radius.
\label{fig:modified-attenuation-function}
}
\end{figure}

Figure \ref{fig:modified-attenuation-function} shows $\fn{F}{Q}$ for the four distinct models of disordered particulate media in $\R^3$: 
overlapping spheres, equilibrium packings, stealthy hyperuniform packings,  and hyperuniform polydisperse packings.
(Its 2D counterpart is provided in the Supplementary Material \cite{SM}.)
We clearly see that these attenuation functions exhibit common large-$Q$ behaviors, regardless of the microstructures.
From the quasistatic to the intermediate-wavelength regimes ($Qa < 1$), however, the attenuation characteristics (imaginary parts $\Im[\fn{F}{Q}]$) 
are considerably different from one model to another.
For example, stealthy hyperuniform media are transparent up to a finite wavelength, and hyperuniform polydisperse packings exhibit much less attenuation than nonhyperuniform systems.

\section{Simulation Procedure to Compute Effective Dynamic Dielectric Constant }
\label{sec:simulation}

In Ref. \cite{Ki20a}, we established preliminary comparisons of strong-contrast approximation \eqref{eq:modified-strong-contrast-approximation_2pt} and numerical simulations via the extended version of the fast-Fourier-Transform-based technique \cite{Mo98,Ey99}.
Because of convergence issues, however, 
in this paper, we employ a more reliable numerical technique, i.e., the FDTD method \cite{Taf13}, using an open-source software package \cite{Os10}.
We focus here on particulate media and take the matrix to be the reference medium (phase 1) and the particles to be the polarized phase (phase 2).

\begin{figure}[h]
\includegraphics[width = 0.5\textwidth]{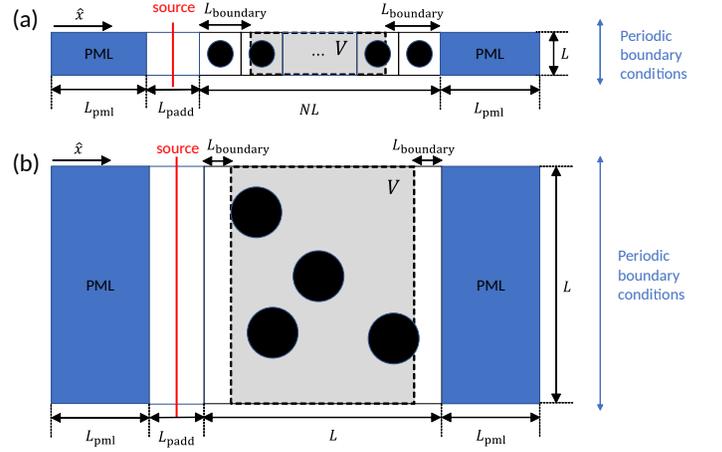}
\caption{
Schematic of the general simulation setup for either  (a) periodic or (b) nonperiodic composites consisting of $N$ spheres of radius $a$
in a matrix.
In both cases, Gaussian pulses of electric fields propagate from the planar sources (shown in red lines) to the packings (shown in black circles).
The wave number (spectrum) of the pulses spans between $\min[k_1]$ and $\max[k_1]$.
Periodic boundary conditions are applied along all directions, except for the propagation direction $\uvect{x}$.
The perfectly matched layers (PML, shown in blue) of thickness $L_\mathrm{PML}$ are placed at both ends of the simulation box to absorb any reflected and transmitted waves. To estimate the effective dielectric constant as described in step 2 below, we consider the subregion $V$ (shown in gray) 
that excludes from the composite within the simulation box two relatively thin slabs of thickness $L_\mathrm{boundary}$ along the propagation direction.
\label{fig:schm-simulations}}
\end{figure}

The general simulation setup is schematically illustrated in Fig. \ref{fig:schm-simulations}.
The simulation procedure for macroscopically isotropic media is carried out in three steps:
\begin{enumerate}
\item For a given medium, we obtain the steady-state spatial distributions of electric field $\fn{E_y}{\vect{r},\omega}$ and electric displacement field $\fn{D_y}{\vect{r},\omega}$ at a given frequency $\omega$ (or the corresponding wave number $k_1$).
Specifically, the planar source generates Gaussian pulses of electric fields that propagate along the $\uvect{x}$ direction
with wave number $k_1$ that spans between $\min[k_1]$ and $\max[k_1]$.
Using the aforementioned MEEP package \cite{Os10}, we compute time evolution of electric field $\fn{E_y}{\vect{r}, t}$ and electric displacement field $\fn{D_y}{\vect{r}, t}$ for a period of time $6\pi \sqrt{\varepsilon_1} / \qty{c \min[k_1]}$, where $c$ is the speed of light in vacuum.
We then compute the temporal Fourier transforms of these fields inside packings.
The values of the simulation parameters (indicated in  Fig. \ref{fig:schm-simulations}) for the
 2D and 3D ordered and disordered models studied in this article are summarized in the Supplementary Material \cite{SM}.

\item  At each value of $k_1$, we postprocess $\fn{E_y}{\vect{r},\omega}$ and $\fn{D_y}{\vect{r},\omega}$ to estimate the effective dielectric constant $\fn{\varepsilon_*}{k_1,\omega}$ of a single configuration by solving the following self-consistent equation:
	\begin{equation}
	\fn{\varepsilon_*}{k_1} = \fn{\overline{\varepsilon_*}}{k_1}, \label{eq:conditions-homogenization}
 	\end{equation}
where
\begin{align*}
	 \fn{\varepsilon_*}{k_1}  \equiv \frac{\fn{\tilde{D}_y}{k_e,\omega}} {\fn{\tilde{E}_y}{k_e, \omega}}, 	
	 \quad
	 \fn{\overline{\varepsilon_*}}{k_1} \equiv \qty(\frac{k_e}{\omega/c})^2	, 
\end{align*}
$k_e$ is a complex-valued effective wave number, and  
\begin{align*}
 \fn{\tilde{D}_y}{q,\omega} \equiv & \frac{1}{\abs{V}} \int_V \fn{D_y}{\vect{r},\omega} e^{-i q\uvect{x}\cdot\vect{r} } \dd{\vect{r}}, \\
 \fn{\tilde{E}_y}{q, \omega} \equiv & \frac{1}{\abs{V}} \int_V \fn{E_y}{\vect{r},\omega} e^{-i q\uvect{x}\cdot\vect{r}} \dd{\vect{r}},
 \end{align*}
where $V$ is a rectangular parallelepiped subregion within the composite (shown in gray in Fig. 6) that is slightly smaller
than the simulation box and is used to reduce undesired boundary effects.
The homogenization task is carried out by numerically finding the minimizer of $\abs{\varepsilon_* - \overline{\varepsilon_*}}^2$ with an initial guess $\overline{\varepsilon_*} = \varepsilon_\mathrm{HS}$ via the Broyden-Fletcher-Goldfarb-Shanno (BFGS) nonlinear optimization algorithm \cite{Liu89}, where $\varepsilon_\mathrm{HS}$ is the Hashin-Shtrikman estimate given by Eq. \eqref{eq:HS-bound}. Details of step 2 are provided in the Supplementary Material \cite{SM}.

\item Steps 1-2 are repeated for a sufficient number of configurations for disordered media.
Then, we compute the effective dielectric constant at a given $k_1$ by ensemble averaging $\fn{\varepsilon_*}{k_1}$, i.e., $\fn{\varepsilon_e}{k_1} = \E{\fn{\varepsilon_*}{k_1}}$.
\end{enumerate}

It is important to note that we ensure that the run times employed in step 1 are sufficiently long 
such that the computed effective dielectric constants achieve stable and accurate steady-state values. 
We emphasize that the result $\fn{\varepsilon_*}{k_1}$ obtained from Eq. \eqref{eq:conditions-homogenization} in step 2 is nonlocal in space because it is calculated from the nonlocal constitutive relation $\fn{\tilde{D}_y}{k_e,\omega} = \fn{\varepsilon_*}{k_1} \fn{\tilde{E}_y}{k_e, \omega} $. 
For periodic media, step 3 is unnecessary because all configurations are identical.

\section{Comparison of Simulations of $\fn{\varepsilon_e}{{\bf k}_1}$ to Various Approximations Formulas}
\label{sim-results}

In this section, we compare our simulations of the effective dynamic dielectric constant $\fn{\varepsilon_e}{{\bf k}_1}$
for various 2D and 3D ordered and disordered model microstructures to the  predictions of the strong-contrast formulas
as well as to conventional approximations, such as MGA \eqref{eq:Maxwell-Garnett_2D}  and QCA
\eqref{eq:quasicrystalline_low-con2}.
Most of these  models provide stringent tests of the predictive power of the  approximations at finite wave numbers because they 
are characterized by nontrivial spatial correlations at intermediate length scales.

\subsection{2D and 3D periodic media}

We first carry out our FDTD simulations for the effective dynamic
dielectric constant $\fn{\varepsilon_e}{\vect{k}_1}$ of 2D and 3D periodic packings
(square and simple-cubic lattice packings), which necessarily
depends on the direction of the incident wave $\vect{k}_1$. While these periodic packings are macroscopically isotropic, due to their cubic symmetry, they are statistically anisotropic
(see Supplementary Material \cite{SM} for details).
For simplicity, we consider only the case where $\vect{k}_1$ is aligned with one of the minimal lattice vectors, i.e., the $\Gamma$-$X$ direction in the first Brillouin zone. 
Such periodic models enable us to validate our simulations because
$\varepsilon_e(\vect{k}_1)$ also can be accurately extracted
from the lowest two photonic bands that are calculated via MPB, an open-source software package \cite{Jo01}.
The results from the band-structure calculations and our FDTD simulations
show excellent agreement. In particular,  our simulations accurately predict two salient dielectric characteristics
that must be exhibited by periodic packings: transparency up to a finite wave number associated with the edge of the first Brillouin zone (i.e., $\Im[\varepsilon_e]=0$ for $0\leq \abs{\vect{k}_1} \lesssim \pi$), and resonancelike attenuation due to Bragg diffraction within the photonic band gap \footnote{Here, the large value of the imaginary part $\Im[\varepsilon_e]$ is due to a small penetration depth of evanescent waves.} (i.e., a peak in $\Im[\varepsilon_e]$ or, equivalently, a sharp transition in $\Re[\varepsilon_e]$ \footnote{The Kramers-Kronig relations \eqref{eq:KKfromImag2Real_temporal} and \eqref{eq:KKfromReal2Imag_temporal} dictate that a resonance phenomenon in the dielectric response, that is, a sharp peak in $\Im[\varepsilon_e]$, must correspond to a sharp transition in $\Re[\varepsilon_e]$, and vice versa.}).
Thus, our numerical homogenization scheme is valid down to intermediate wavelengths
(see the Supplementary Material \cite{SM} for comparison of the band-structure and FDTD computations).

\begin{figure}[h!]
\subfloat[]{
\includegraphics[width = 0.42\textwidth]{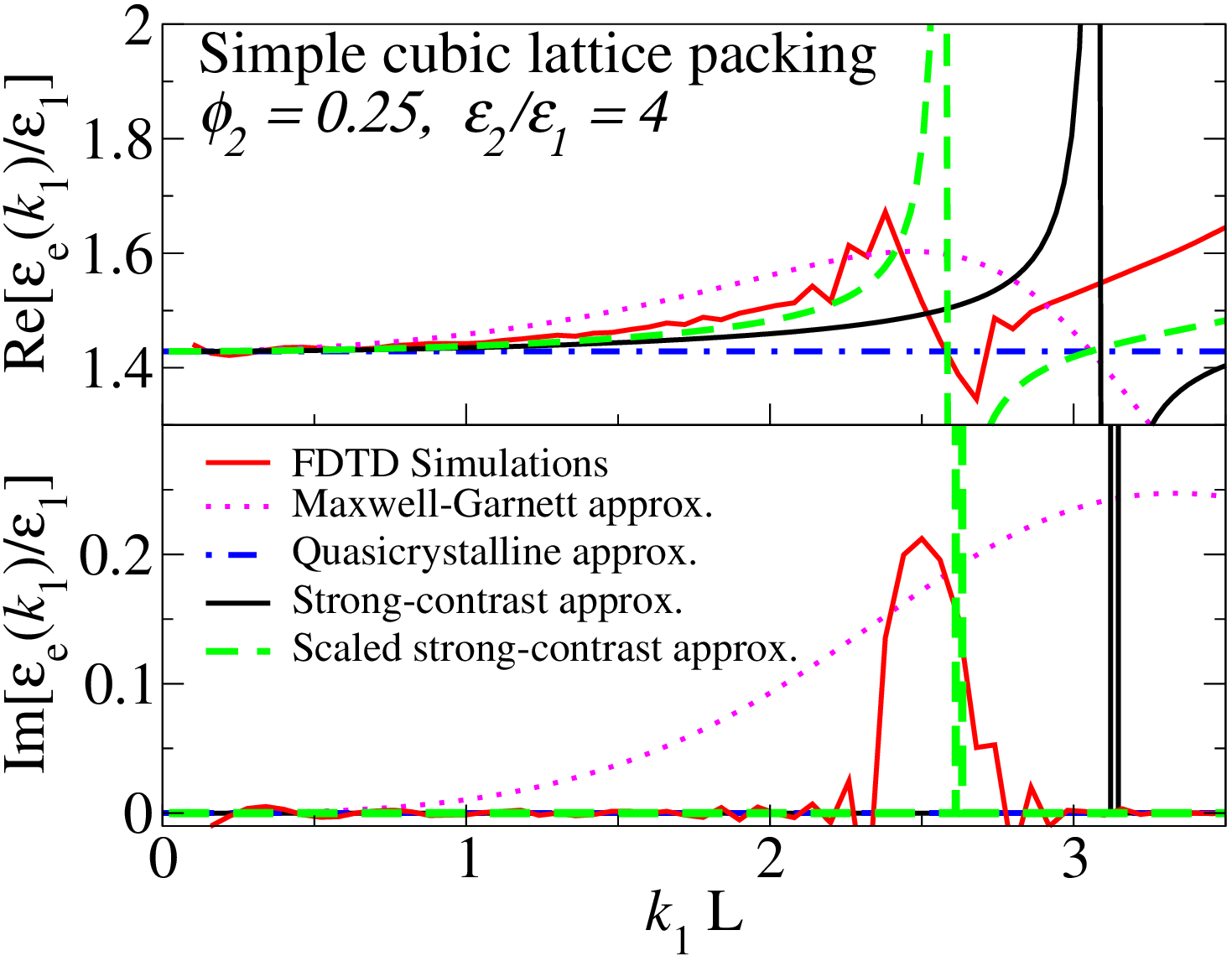}}

\subfloat[]{
\includegraphics[width = 0.42\textwidth]{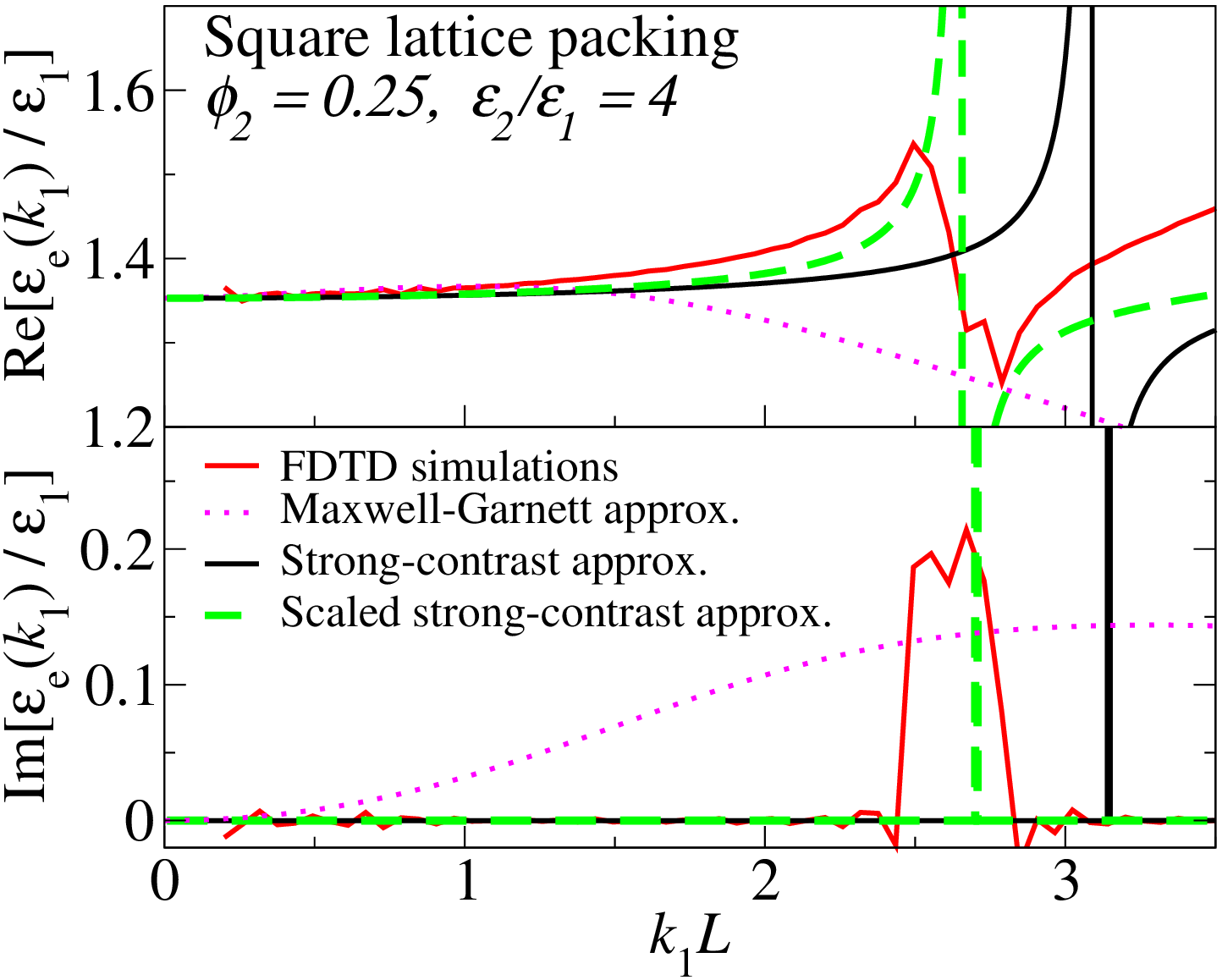}}
\caption{Comparison of the predictions of the strong-contrast  formulas, Eqs. \eqref{eq:modified-strong-contrast-approximation_2pt} and \eqref{eq:scaled-2pt-approx}, 
to the Maxwell-Garnett [Eqs. \eqref{eq:Maxwell-Garnett_2D} and \eqref{eq:Maxwell-Garnett_3D}] and QCA
\eqref{eq:quasicrystalline_low-con2} approximations for the effective dynamic dielectric constant $\fn{\varepsilon_e}{\vect{ k}_1}$ as a function of the dimensionless wave number $k_1L$ of periodic packings to our corresponding
computer simulation results.
We consider (a) 3D simple cubic lattice and (b) 2D square lattice of packing fraction $\phi_2=0.25$ and contrast ratio $\varepsilon_2/\varepsilon_1 = 4$.
    Here, $k_1$ is the wave number in the reference (matrix) phase  along the $\Gamma$-$X$ direction, and $L$ is the side length of a unit cell.}
\label{fig:periodic-simulations}
\end{figure}

Importantly, while our strong-contrast approximations Eqs. \eqref{eq:modified-strong-contrast-approximation_2pt} and \eqref{eq:scaled-2pt-approx}
account for directionality of the incident waves, the  MGA and QCA are independent of the direction of ${\bf k}_1$.
In Fig. \ref{fig:periodic-simulations}, the FDTD simulation results are
compared with the MGAs for $d=2,3$ [Eqs. \eqref{eq:Maxwell-Garnett_2D} and \eqref{eq:Maxwell-Garnett_3D}],
QCA \eqref{eq:quasicrystalline_low-con2} for $d=3$, as well as the unscaled and scaled strong-contrast
approximations \eqref{eq:modified-strong-contrast-approximation_2pt} and \eqref{eq:scaled-2pt-approx} for $d=2,3$. While all approximations agree with the FDTD simulations in the quasistatic regime, the MGA and QCA fail to capture properly two key features: no loss of energy up to a finite wave number and resonancelike attenuation in the band gaps.
Each strong-contrast approximation captures both of these salient characteristics. However, it
is noteworthy that the scaled strong-contrast approximation [Eq. \eqref{eq:scaled-2pt-approx}] agrees very well with the FDTD simulations.
For contrast ratios $\varepsilon_2/\varepsilon_1 <1$,  FDTD simulations are also in very good agreement with the predictions of strong-contrast approximations for a wide range of wave numbers, as detailed in the Supplementary Material \cite{SM}.

\subsection{Disordered nonhyperuniform and hyperuniform media}

\begin{figure}[h!]
\subfloat[]{\includegraphics[width=0.42\textwidth]{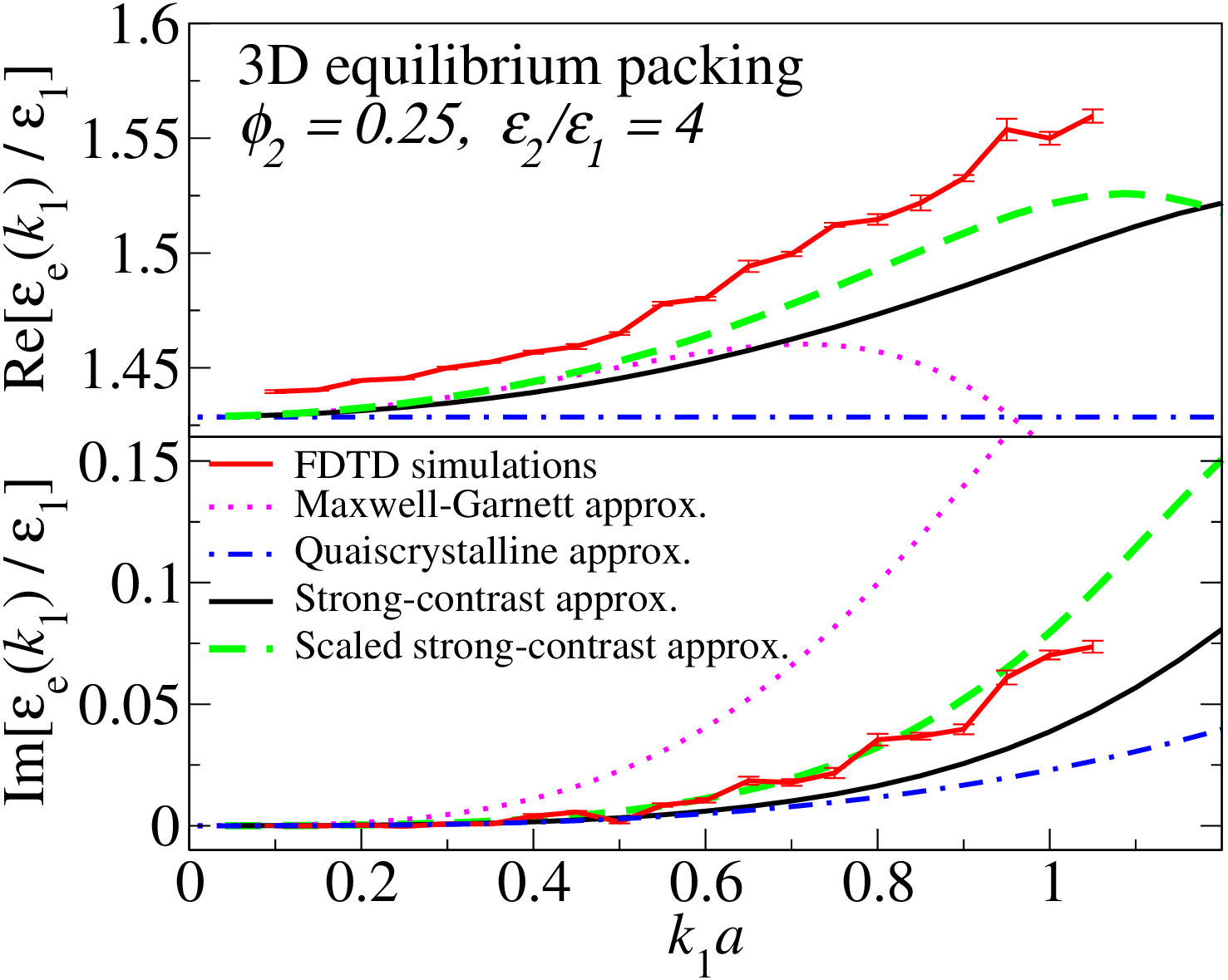}}

\subfloat[]{\includegraphics[width=0.42\textwidth]{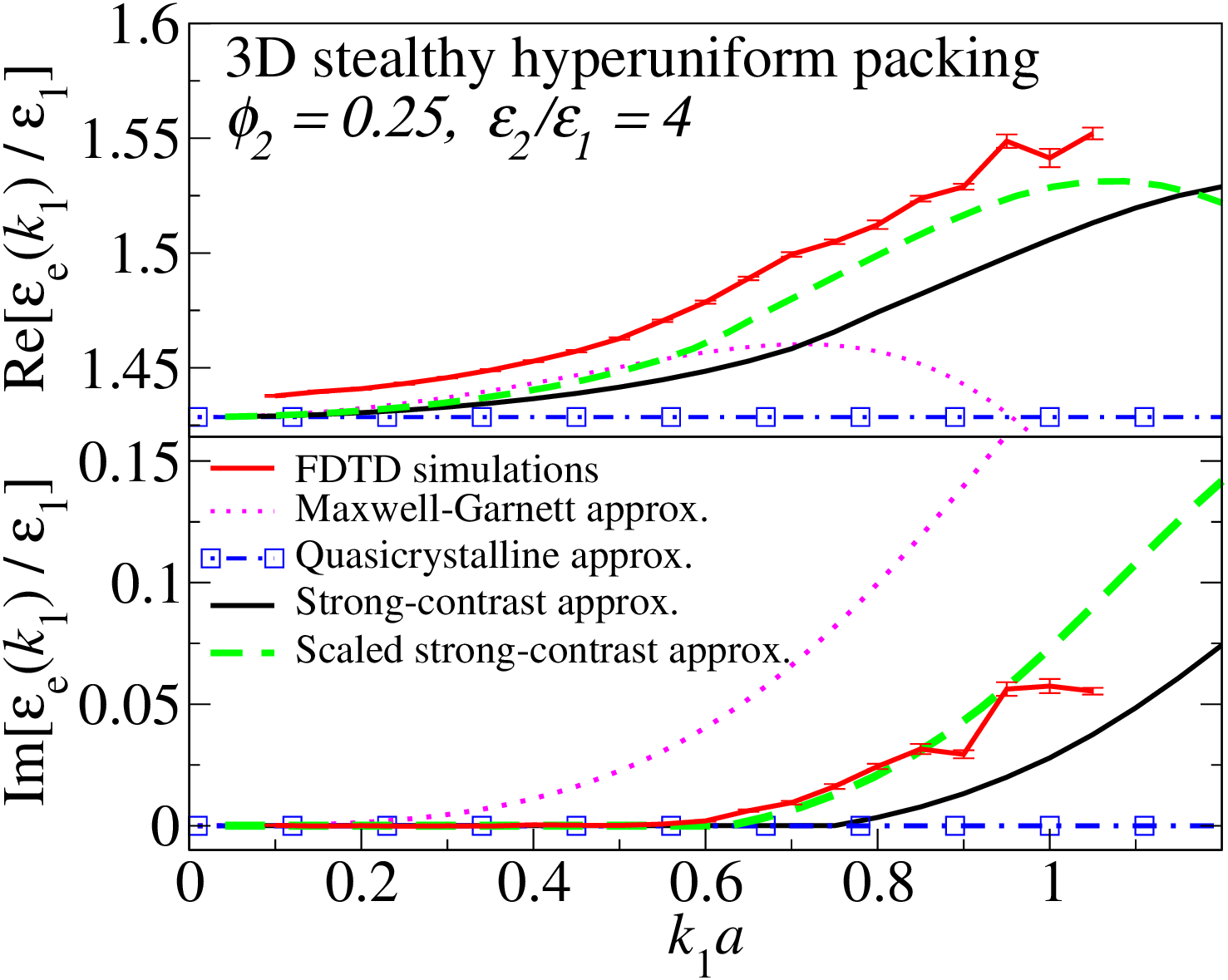}}
\caption{
Comparison of the predictions of the strong-contrast formulas Eqs. \eqref{eq:modified-strong-contrast-approximation_2pt} and \eqref{eq:scaled-2pt-approx} to the MGA\eqref{eq:Maxwell-Garnett_3D} and QCA
\eqref{eq:quasicrystalline_low-con2} approximations for the effective dynamic dielectric constant $\fn{\varepsilon_e}{ k_1}$ as a function of the dimensionless wave number $k_1a$ of 3D disordered sphere packings to our corresponding
computer simulation results.
We consider (a) equilibrium packings and (b) stealthy hyperuniform packings [$\spD{Q}=0$ for $0 \le Qa<1.5$] of sphere radius $a$, packing fraction $\phi_2 = 0.25$, and phase-contrast ratio $\varepsilon_2 / \varepsilon_1 = 4$.
    Here, $k_1$ is the wave number in the reference (matrix) phase, and the error bars in the FDTD simulations represent the standard errors over independent configurations.}
    \label{fig:3D-disordered-simulations}
\end{figure}

To test the predictive capacity of approximation formulas for $\varepsilon_e(k_1)$ for
disordered media as measured against simulations, we choose to study
two distinctly different models: disordered nonhyperuniform packings (equilibrium packings)  and disordered stealthy hyperuniform disordered packings 
[i.e., $\spD{Q}=0$ for $0 \le Qa< 1.5$] for both 2D and 3D. 
Again, we compare  our simulation results to the MGAs [Eqs. \eqref{eq:Maxwell-Garnett_2D} and \eqref{eq:Maxwell-Garnett_3D}], QCA \eqref{eq:quasicrystalline_low-con2}, strong-contrast approximation \eqref{eq:modified-strong-contrast-approximation_2pt}, and the scaled counterpart \eqref{eq:scaled-2pt-approx}.
The conventional approximations fail to capture spatial dispersion effects. 
Specifically, the MGA  neglects any microstructural information, except for the particle shape, and thus cannot
account for long-range correlations, such as the lossless property of stealthy hyperuniform media.
By contrast, while the QCA formula yields better estimates of $\Im[\varepsilon_e]$ for nonhyperuniform systems,
it cannot generally  capture the correct transparency characteristics of hyperuniform systems, e.g.,
it incorrectly predicts $\Im[\fn{\varepsilon_e}{k_1}]=0$ for all wave numbers, regardless of whether the medium 
is stealthy hyperuniform or nonstealthy hyperuniform; see Fig. \ref{fig:3D-disordered-simulations} (b).

On the other hand, the scaled strong-contrast approximation provides excellent  estimates of $\varepsilon_e(k_1)$ for both disordered models, even for large wave numbers
($0 \le k_1 a \le 1$);
see Fig. \ref{fig:3D-disordered-simulations}. Moreover, the predictions of both strong-contrast approximations accurately capture the salient microstructural
differences between the nonhyperuniform and hyperuniform models because they incorporate spatial correlations at finite wavelengths
via the spectral density $\spD{Q}$. For example, they properly predict that stealthy hyperuniform media are lossless up to a finite wave number,
even if at different cutoff values; see Eq. \eqref{eq:transparency-regime-strong-contrast}.
Corresponding 2D results are presented in the Supplementary Material \cite{SM} because they are qualitatively the same as the 3D results.

\section{Predictions of Strong-Contrast Approximations for Disordered Particulate Media}\label{predictions}

\begin{figure}[h!]
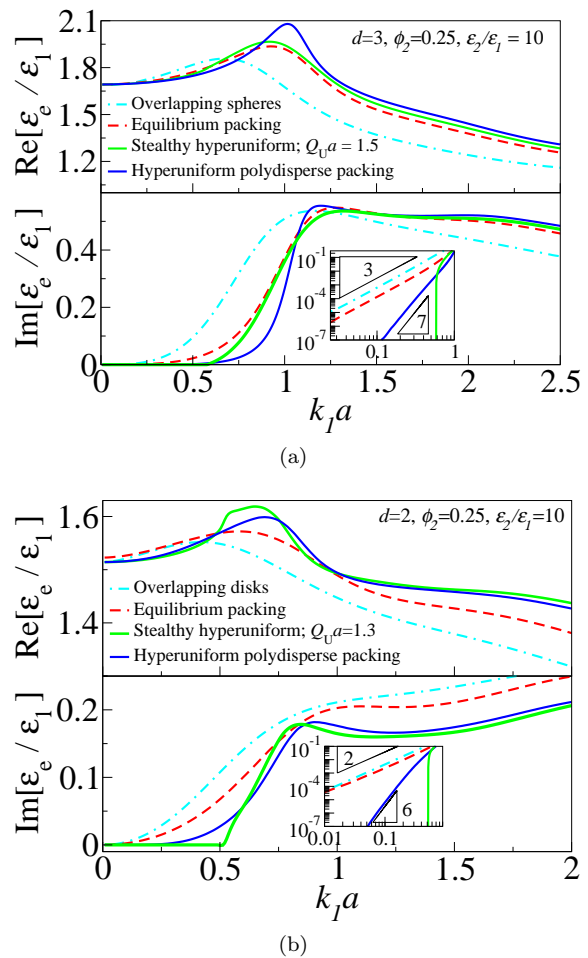

\subfloat[]{
\includegraphics[width = 0.42\textwidth]{fig10a.eps}}

\subfloat[]{
\includegraphics[width = 0.42\textwidth]{fig10b.eps}}
\caption{Predictions of the scaled strong-contrast approximation \eqref{eq:scaled-2pt-approx} for the effective dynamic dielectric constant $\varepsilon_e(k_1)$ as a function of the dimensionless wave number $k_1a$ of the four models of disordered media at volume fraction $\phi_2 = 0.25$ and contrast ratio $\varepsilon_2 /\varepsilon_1 = 10$: (a) three dimensions and (b) two dimensions.
The inset in the lower panel is the log-log plot of the larger panel.} 
\label{fig:epsilons-vs-k}
\end{figure}

Having established the accuracy of the scaled strong-contrast approximation \eqref{eq:scaled-2pt-approx} for ordered
and disordered media in the previous section, we now apply it to the four different disordered models discussed in Sec. \ref{sec:models} in order to study
how $\fn{\varepsilon_e}{k_1}$ varies with the microstructure.
We first study how $\fn{\varepsilon_e}{k_1}$ varies with $k_1$ at a fixed contrast ratio $\varepsilon_2/\varepsilon_1=10$ for the four models; see Fig. \ref{fig:epsilons-vs-k}.
According to Eq. (\ref{eq:Im-diel-quasistatic}), nonhyperuniform and hyperuniform media in the quasistatic regime
have the different scalings, i.e., $\Im[\fn{\varepsilon_e}{k_1}] \sim {k_1}^d$ and $\Im[\fn{\varepsilon_e}{k_1}] \sim {k_1}^{d+\alpha}$, respectively, where $\alpha >0$ for hyperuniform systems. This implies that  hyperuniform media are less lossy than their nonhyperuniform counterparts as $k_1$ tends to zero, as seen in the insets of Fig. \ref{fig:epsilons-vs-k}.
Moreover, beyond the quasistatic regime, each model exhibits ``effective" transparency for a range of wave numbers that depends on the microstructure.
For 2D and 3D models, hyperuniform polydisperse packings tend to be effectively transparent for a wide range of wave numbers compared to the nonhyperuniform ones, while the stealthy hyperuniform systems are perfectly transparent for the widest range of wave numbers, as established in Eq. \eqref{eq:transparency-regime-strong-contrast} and Sec. \ref{sim-results}.
For each model, the ``effective" transparency spectral range must be accompanied by {\it normal dispersion} [i.e., an increase in $\Re[\fn{\varepsilon_e}{k_1}]$ with $k_1$] \cite{Ag84} because our strong-contrast approximation is consistent with the Kramers-Kronig relations (see Appendix \ref{sec:KK}).
Moreover, we  see that {\it anomalous dispersion} [i.e., a decrease in $\Re[\fn{\varepsilon_e}{k_1}]$ with $k_1$]
occurs at wave numbers larger but near the respective transition between the effective transparency and appreciable attenuation, which again is dictated by the Kramers-Kronig relations.
The specific anomalous dispersion behavior is microstructure dependent.

\begin{figure}[h!]
\subfloat[]{
\includegraphics[width = 0.42\textwidth]{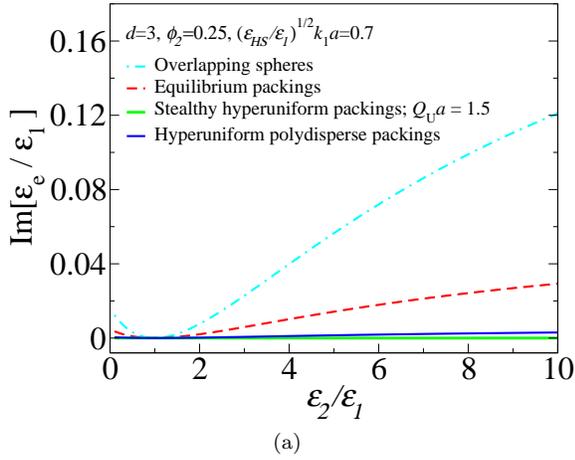}}

\subfloat[]{
\includegraphics[width = 0.42\textwidth]{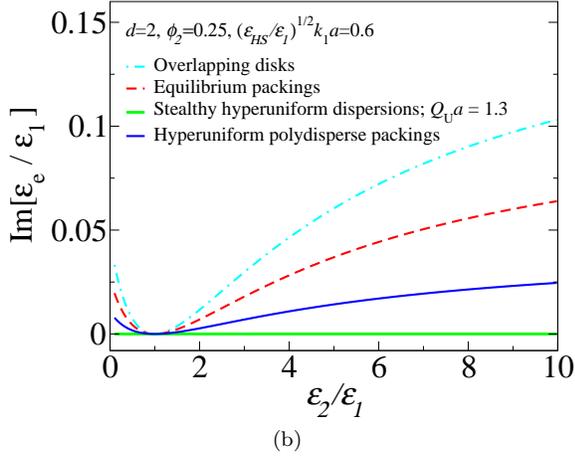}}
\caption{
Predictions of the strong-contrast approximation \eqref{eq:scaled-2pt-approx} for the effective dynamic dielectric constant $\varepsilon_e(k_1)$  as function of dielectric-contrast ratio $\varepsilon_2 /\varepsilon_1$ for the four disordered models, as per Fig. \ref{fig:epsilons-vs-k}, at packing fraction $\phi_2 = 0.25$ and wave number (a) $\sqrt{\varepsilon_\mathrm{HS}/\varepsilon_1} ~k_1 a = 0.7$ in three dimensions and (b) $\sqrt{\varepsilon_\mathrm{HS}/\varepsilon_1} ~k_1 a = 0.6$ in two dimensions.
\label{fig:epsilon-vs-contrast_3D}}
\end{figure}

We now examine how the imaginary part $\Im[\varepsilon_e]$ varies with the contrast ratio $\varepsilon_2/\varepsilon_1$ 
for the disordered models for a given large wave number $k_1$ inside the transparency interval for 2D and 3D stealthy hyperuniform systems.
These results are summarized in Fig. \ref{fig:epsilon-vs-contrast_3D}.
The disparity in the attenuation characteristics across microstructures widens significantly as the contrast ratio increases.
Clearly, overlapping spheres are the lossiest systems.
Hyperuniform polydisperse packings can be nearly as lossless as stealthy hyperuniform ones.

\begin{figure}
\begin{center}
\includegraphics[width = 0.4\textwidth]{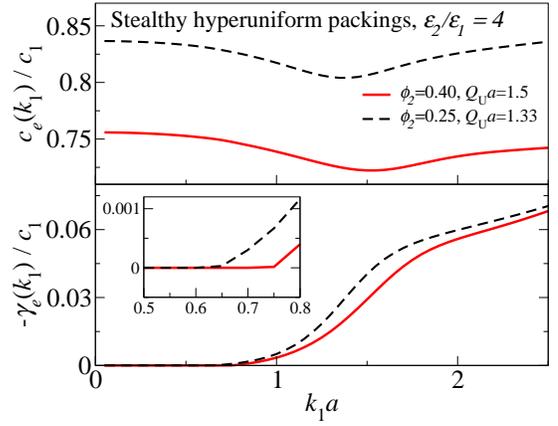}
\end{center}
\caption{
Predictions of the scaled strong-contrast approximation \eqref{eq:scaled-2pt-approx} for the effective wave speed $c_e$ and the negative
of the attenuation coefficient $\gamma_e$ as a function of the dimensionless wave number $k_1a$ for 3D stealthy hyperuniform sphere packings of contrast ratio $\varepsilon_2/\varepsilon_1 = 4$
at two different packing fractions: $\phi_2=0.4$ with $Q_\mathrm{U}a = 1.5$ and $\phi_2=0.25$ with $Q_\mathrm{U}a\approx 1.33$.
The inset is a magnification of the lower panel.}
\label{fig:3D-SHU-phis}
\end{figure}

We also study the effect of packing fraction $\phi_2$ on  the effective phase speed $\fn{c_e}{k_1}$ and effective attenuation coefficient $\fn{\gamma_e}{k_1}$,
as defined by Eqs. \eqref{speed} and \eqref{atten}, respectively.
For concreteness, we focus on  3D stealthy hyperuniform packings. We first generate
such packings at a packing fraction $\phi_2=0.4$ and $Q_\mathrm{U}a = 1.5$, as described in Sec. \ref{stealth}.
Without changing particle positions, we then shrink  particle radii to attain a packing fraction $\phi_2=0.25$, whose stealthy regime is now $Q_\mathrm{U}a \approx 1.33$.
The coefficients $\fn{c_e}{k_1}$ and $\fn{\gamma_e}{k_1}$ for these packings with $\varepsilon_2/\varepsilon_1=4$ are estimated from the scaled 
 approximation \eqref{eq:scaled-2pt-approx}; see Fig. \ref{fig:3D-SHU-phis}.
It is seen that the waves propagate significantly more slowly through the denser medium due to an increase in multiple-scattering events.
Moreover, the transparency intervals (wave-number ranges where the effective attenuation coefficients vanish) are larger for the packing with the higher stealthy cutoff value $Q_U a=1.5a$ ($\phi_2=0.4$),
as predicted by Eq. \eqref{eq:transparency-regime-strong-contrast}.

\section{Discussion}

All previous closed-form homogenization estimates of the effective dynamic dielectric constant apply only at long wavelengths (quasistatic regime) and for very special  {\it macroscopically isotropic} disordered composite microstructures, namely, nonoverlapping spheres or spheroids in a matrix. 
In this work, we lay the theoretical foundation that enables us to substantially extend previous
work in both its generality and applicability.
First, we derive exact  homogenized constitutive relations for the effective dynamic dielectric constant tensor $\tens{\varepsilon}_e(\vect{k}_q)$  that are {\it nonlocal in space} from first principles.
Second, our strong-contrast representation of $\tens{\varepsilon}_e(\vect{k}_q)$ exactly accounts for complete microstructural information (infinite set of $n$-point correlation functions) for {\it arbitrary microstructures} and hence multiple scattering  
to all orders for the range of wave numbers for which our extended homogenization theory applies, i.e.,  $0 \le  |\vect{k}_q| \ell \lesssim 1$
(where $\ell$ is a characteristic heterogeneity length scale).
Third, we extract from the exact expansions accurate nonlocal closed-form approximate formulas for  $\tens{\varepsilon}_e(\vect{k}_q)$, relations  \eqref{eq:modified-strong-contrast-approximation_2pt} and \eqref{eq:scaled-2pt-approx}, which are resummed representations
of the exact expansions  that incorporate microstructural information through the spectral density $\spD{\vect{Q}}$, which is easily ascertained for general microstructures either 
theoretically, computationally, or via scattering experiments. Depending on whether the high-dielectric phase percolates or not, the wide
class of microstructures that we can treat includes
particulate media consisting of identical or polydisperse particles of general shape (ellipsoids, cubes, cylinders, polyhedra)
with prescribed orientations that may or not {\it overlap}, and cellular networks as well as media without well-defined inclusions (Sec. \ref{trunc}). 
Our approximations account for multiple scattering across a range of wave numbers. 
Fourth, we carry out precise full-waveform simulations for various  2D and 3D models of ordered and disordered media
to validate the accuracy of our nonlocal microstructure-dependent approximations for wave numbers well beyond the quasistatic regime.

%

Having established the accuracy of the scaled strong-contrast approximation \eqref{eq:scaled-2pt-approx}, we then apply it to four models of 2D and 3D disordered media (both nonhyperuniform and hyperuniform)
to investigate the effect of microstructure on the effective wave characteristics. Among other findings, we show that disordered hyperuniform media are generally less lossy than their nonhyperuniform counterparts. We also find that our {\it scaled}  formula \eqref{eq:scaled-2pt-approx}
accurately predicts that disordered stealthy hyperuniform media possess a transparency wave number interval 
$[0,0.5\,Q_\mathrm{U} (\varepsilon_\mathrm{HS}/\varepsilon_q)^{-1/2})$ [cf. \eqref{eq:transparency-regime-strong-contrast}], where most 
nonhyperuniform disordered media are opaque.
Note that, using multiple-scattering simulations, Leseur \textit{et al.} \cite{Le16} were the first to show  that stealthy hyperuniform systems should exhibit a transparency interval, but for ``point" scatterers, not the finite-sized
scatterers considered here. Interestingly, their transparency-interval prediction coincides
with the one predicted by our less accurate strong-contrast formula  [cf. Eq. \eqref{eq:transparency-regime-strong-contrast}].

The accuracy of our nonlocal closed-form formulas has important practical implications, since one can now use them to accurately and efficiently predict the effective wave characteristics well beyond the quasistatic regime of a wide class of composite microstructures without having to perform computationally expensive full-blown simulations. 
Thus, our nonlocal formulas can be used to accelerate the discovery of novel electromagnetic composites by appropriate tailoring of the spectral densities 
and then constructing the corresponding microstructures by using the Fourier-space inverse methods \cite{Ch18a}.
For example, from our findings in the present study, it is clear that stealthy disordered particulate media can be employed as low-pass filters that transmit waves
``isotropically” up to a selected wave number. Moreover, using the  spectral densities
of the type found by Chen and Torquato \cite{Ch18a} for stealthy hyperuniform packings
(characterized by a peak value at $Q=Q_\text{U}$ with intensities that rapidly decay to zero for larger wave numbers)
and formula \eqref{eq:scaled-2pt-approx},
one can design materials with  refractive indices that abruptly change over a narrow range
of wave numbers. Of course, one could also explore the design space of effective 
wave properties of nonhyperuniform
disordered composite media for potential applications.

Previously, disordered media were often described using cluster expansions \cite{Fr68,Sh95,Ts01}, while ordered media were often studied through dispersion relations and band-structures calculations.
Accordingly, it has been of primary importance to bridge the gap between the treatments of ordered and disordered to better understand the optical properties of correlated media.
Thus, our work represents an initial step toward a unified theory to describe the effective optical properties of both ordered and disordered microstructures over a wide range of incident wavelengths.

There are a variety of directions for future research.
First, our formalism can be straightforwardly
extended to heterogeneous materials composed of more than two phases or continuous media.
Second, it is also of interest to extend our formalism to applications and phenomena (e.g., magnetic effects) relevant to the smaller wavelengths noted for metamaterials \cite{li04, sm04, wa12}.


{\appendix

\section{Different Expansions as a Result of Different Exclusion-Region Shapes} \label{app:exclusion-regions}

To get a sense of how the resulting expansions change due to the choice of the exclusion-region shape, we 
consider the aforementioned oriented spheroidal exclusion region in the two limiting disk-like and needle-like cases.
Comparing the expansion parameters in the limit cases given in Eq. \eqref{eq:exclusion-region} to the strong-contrast expansion with a spherical exclusion region given in Eq. \eqref{isotropic-strong}, one can obtain the counterparts of Eq. \eqref{isotropic-strong} with disk-like and needle-like exclusion regions.
\begin{widetext}
Specifically, we replace parameters in Eq. \eqref{isotropic-strong} according to the following mappings:
\begin{align*}
\beta_{pq} \to&
(\varepsilon_p-\varepsilon_q) / (d \varepsilon_p),&
\frac{\fn{{\varepsilon}_e}{\vect{k}_q} + (d-1)\varepsilon_q{}}{\fn{{\varepsilon}_e}{\vect{k}_q} - \varepsilon_q {}} 
\to & \frac{d\fn{{\varepsilon}_e}{\vect{k}_q}}{\fn{{\varepsilon}_e}{\vect{k}_q} -\varepsilon_q{}},& ~~~~
\text{disk-like}, \\
\beta_{pq} \to&
(\varepsilon_p-\varepsilon_q)/(d\varepsilon_q),&
\frac{\fn{{\varepsilon}_e}{\vect{k}_q} + (d-1)\varepsilon_q{}}{\fn{{\varepsilon}_e}{\vect{k}_q} - \varepsilon_q{}} 
\to & \frac{d\varepsilon_q{}}{\fn{{\varepsilon}_e}{\vect{k}_q}-\varepsilon_q {}},& ~~~~
\text{needle-like},
\end{align*}
resulting in the following expansions, respectively,
\begin{align}
{\phi_p}^2 \qty(\frac{\varepsilon_p-\varepsilon_q}{d\varepsilon_p})^2
\frac{d\fn{{\varepsilon}_e}{\vect{k}_q}}{\fn{{\varepsilon}_e}{\vect{k}_q} -\varepsilon_q{}}
=& \phi_p \qty(\frac{\varepsilon_p-\varepsilon_q}{d\varepsilon_p}){} - \sum_{n=2}^\infty
\fn{{A}_n^{(p)}}{\vect{k}_q;A^*=1} \qty(\frac{\varepsilon_p-\varepsilon_q}{d\varepsilon_p})^{n}, 
\label{strong-disk-like}\\
{\phi_p}^2 \qty(\frac{\varepsilon_p-\varepsilon_q}{d\varepsilon_q})^2
\frac{d\varepsilon_q{}}{\fn{{\varepsilon}_e}{\vect{k}_q}-\varepsilon_q {}}
=& \phi_p \qty(\frac{\varepsilon_p-\varepsilon_q}{d\varepsilon_q}){} - \sum_{n=2}^\infty
\fn{{A}_n^{(p)}}{\vect{k}_q;A^*=0} \qty(\frac{\varepsilon_p-\varepsilon_q}{d\varepsilon_q})^{n}.
\label{strong-needle-like}
\end{align}
Here the functionals $\fn{\tens{A}_n^{(p)}}{\vect{k}_q;A^*}$ are identical to Eqs. \eqref{eq:A2-tensor_direct-space} and \eqref{eq:An-tensor_direct-space}, except for the  exclusion-region shape.

In the Supplementary Material \cite{SM}, we discuss how to obtain the analogs of Eqs. (A1) and (A2) that apply to macroscopically anisotropic media. 
The corresponding series expansions involve tensorial expansion parameters that have rapid convergence properties for stratified and transversely isotropic media, respectively.


\end{widetext}
	
\section{Kramers-Kronig Relations}\label{sec:KK}

Kramers-Kronig relations connect the real and imaginary parts of any complex function that is analytic in the upper half-plane and meets
mild conditions \cite{Ja90,Mi97}.
Since causality in a dielectric response function of a homogeneous material implies such analyticity properties, the Kramers-Kronig
relations enable one to directly link the real part of a response function 
to its imaginary part or vice versa, even if the real or imaginary parts are 
only available in a finite frequency range \cite{Ja90,Mi97}. 
Thus, when a heterogeneous
material can be treated as a homogeneous material with a  dynamic effective dielectric constant $\varepsilon_e(k_q)$,
Kramers-Kronig relations immediately apply to  the exact strong-contrast
expansion \eqref{eq:expansions_remainder}, i.e.,
\begin{align}
\Re[\fn{\varepsilon_e}{k_q}] = & \fn{\varepsilon_e}{\infty} + \frac{2}{\pi} \mathrm{p.v.} \int_0^\infty \dd{q} \frac{q \Im[\fn{\varepsilon_e}{q}]}{q^2 - {k_q}^2},  \label{eq:KKfromImag2Real_temporal} \\
\Im[\fn{\varepsilon_e}{k_q}] = & - \frac{2k_q}{\pi} \mathrm{p.v.} \int_0^\infty \dd{q} \frac{ \Re[\fn{\varepsilon_e}{q}]-\fn{\varepsilon_e}{\infty}}{q^2 - {k_q}^2},	\label{eq:KKfromReal2Imag_temporal}
\end{align}
where we assume a linear dispersion relation in the reference phase (i.e., $k_q = \sqrt{\varepsilon_q}\omega/c$) and 
$\lim_{\omega \to \infty } \fn{\varepsilon_e}{\omega}=\varepsilon_q$ is real-valued.
The Kramers-Kronig relations may or may not
be obeyed when the strong-contrast expansion is truncated at the two-point level, yielding Eq. (\ref{eq:modified-strong-contrast-approximation_2pt}).  
Here we analytically  show that the effective dielectric constant $\fn{\varepsilon_e}{k_q}$ for isotropic media obtained 
from either the unscaled or scaled strong-contrast formulas [see Eqs.\eqref{eq:modified-strong-contrast-approximation_2pt} and \eqref{eq:scaled-2pt-approx}] also 
satisfies the Kramers-Kronig relations.

We begin by rewriting either
 strong-contrast approximation as $\fn{\varepsilon_e}{k_q} \approx \varepsilon_q+\qty[a+b\fn{F}{k_q}]^{-1} $, where $a$ and $b$ are nonzero real numbers.
The general analytic properties of the nonlocal attenuation 
function $\fn{F}{Q}$ (detailed in the Supplementary Material \cite{SM})  induce  $\varepsilon_e({k_q})$ to have
the following  three properties necessary to satisfy the Kramers-Kronig relations \eqref{eq:KKfromImag2Real_temporal} and \eqref{eq:KKfromReal2Imag_temporal}: 
(i) $\varepsilon_e(k_q)$ is an analytic function in the \textit{upper half-plane} of $k_q$, (ii) $\fn{\varepsilon}{k_q} -\varepsilon_q$ 
vanishes like $1/\abs{k_q}$ as $\abs{k_q}$ goes to infinity,
and  (iii) $\Re[\fn{\varepsilon_e}{k_q}]$ and $\Im[\fn{\varepsilon}{k_q}]$ are even and odd functions of $k_q$, respectively.
Property (i) is valid if $a+b\fn{F}{k_q}\neq 0$, which is met for all disordered systems considered here.
The fact that the strong-contrast approximations satisfy Eqs. \eqref{eq:KKfromImag2Real_temporal} and \eqref{eq:KKfromReal2Imag_temporal}  makes physical sense 
since  $\fn{F}{Q}$ involves $\fn{\tens{G}^{(q)}}{\vect{x},\vect{x}'}$ [cf. Eq. \eqref{eq:relation_GreenFn_D_H}], which  is the temporal Fourier transform of the \textit{retarded} Green's function $\fn{\tens{G}^{(q)}}{\vect{x},t,\vect{x}',t'}$ \cite{Ja90} that accounts for causality.
We also numerically show in the Supplementary Material \cite{SM} that our approximations obey Eqs. (\ref{eq:KKfromImag2Real_temporal}) and (\ref{eq:KKfromReal2Imag_temporal}).

\section{Strong-Contrast Approximation at the Three-Point Level}\label{3-pt} 

\begin{widetext}
Here we present the strong-contrast approximation at the three-point level for a spherical exclusion region.
It is obtained from Eq. \eqref{isotropic-strong} by setting $A_n^{(p)} = 0$ for $n\geq 4$ and by solving it in $\fn{\varepsilon_e}{\vect{k}_q}$:
\begin{align}
\frac{\fn{\varepsilon_e}{\vect{k}_q}}{\varepsilon_q} 
&= 
1+ \frac{d{\beta_{pq}} {\phi_p} ^2}{\phi_p(1-\beta_{pq} \phi_p) + (d-1)\pi /[2^{d/2}\fn{\Gamma}{d/2}]\beta_{pq} \fn{F}{\vect{k}_q}  -{\beta_{pq}}^2 \fn{A_3^{(p)}}{\vect{k}_q} },
\end{align}
where the local three-point parameter is given as

\begin{align}
\fn{A_3^{(p)}}{\vect{k}_q} \equiv & -\frac{(d\varepsilon_q)^2}{\phi_p}
	\int_\epsilon \dd{\vect{x}_1}\dd{\vect{x}_2}
	\frac{1}{d}\Tr[\fn{\tens{H}^{(q)}}{\vect{x}_1-\vect{x}_2} e^{-i\vect{k}_q\cdot(\vect{x}_1-\vect{x}_2)}\cdot
	\fn{\tens{H}^{(q)}}{\vect{x}_2-\vect{x}_3} e^{-i\vect{k}_q\cdot(\vect{x}_2-\vect{x}_3)}]
	\nonumber \\
	&\times \fn{\Delta_3^{(p)}}{\vect{x}_1,\vect{x}_2,\vect{x}_3} \label{eq:nonlocal-A3-direct} \\
	=&-\frac{1}{\phi_p (2\pi)^{2d}} \int \dd{\vect{q}_1}\dd{\vect{q}_2} \frac{1}{{q_1}^2-{k_q}^2}\frac{1}{{q_2}^2-{k_q}^2} \Bigg\{ (d-1)^2{k_q}^4 + {q_1}^2{q_2}^2 \qty[d(\uvect{q}_1 \cdot\uvect{q}_2)^2-1]\Bigg\} \nonumber \\
	&\times \fn{\tilde{\Delta}_3^{(p)}}{\vect{q}_1+\vect{k}_q,\vect{q}_2+\vect{k}_q},  \label{eq:nonlocal-A3-Fourier}
\end{align}
where, due to the statistical homogeneity, 
$$
\fn{\tilde{\Delta}_3^{(p)}}{\vect{q}_1,\vect{q}_2} \equiv \int \dd{\vect{r}_1}\dd{\vect{r}_2} e^{-i\vect{q}_1\cdot\vect{r}_1} e^{-i\vect{q}_2\cdot\vect{r}_2} 
\qty[\fn{S_2^{(p)}}{\vect{r}_1}\fn{S_2^{(p)}}{\vect{r}_2} -\phi_p \fn{S_3^{(p)}}{\vect{r}_1,\vect{r}_2}].
$$
Note that Eq. \eqref{eq:nonlocal-A3-Fourier} is obtained from Eq. \eqref{eq:nonlocal-A3-direct} via Parseval's theorem.
The static limit of Eq. \eqref{eq:nonlocal-A3-direct} for statistically isotropic media is given by \cite{To02a}
\begin{align}
\fn{A_3^{(p)}}{0} 
=& (d-1)\phi_p(1-\phi_p) \zeta_p =  \qty(\frac{d}{\Omega_d})^2\iint \frac{\dd{\vect{r}}}{r^d}\frac{\dd{\vect{s}}}{s^d} [d(\uvect{r}\cdot\uvect{s})^2-1]\qty[\fn{S_3^{(p)}}{r,s,t}-\fn{S_2^{(p)}}{r}\fn{S_2^{(p)}}{s}/\phi_p] ,
\end{align}
where $\Omega_d$ is the surface area of a unit sphere in $\mathbb{R}^d$, $t\equiv\abs{\vect{r}-\vect{s}}$ and
the parameter $\zeta_p$ lies in the closed interval $[0,1]$.

\end{widetext}

\section{Strong-Contrast Formula for Phase-Inversion Symmetric Media}

To further illustrate the flexibility and power of the strong-contrast formalism,
we apply it to treat  media with phase-inversion symmetry.
A two-phase medium possesses {\em phase-inversion symmetry}  if
the morphology of phase 1 at volume fraction $\phi_1$
is statistically identical to that of phase 2 in the system
when the volume fraction of phase 1 is $1-\phi_1$ \cite{To02a}.
In particular, we follow the same procedure used by Torquato
\cite{To02a} to derive strong-contrast expansions designed to
apply to media with phase-inversion symmetry in the static limit 
and which can be regarded to be expansions that perturb around the microstructures
corresponding to the ``self-consistent" formula.
All we need to do is add the expansion \eqref{isotropic-strong} with $p=2$ and $q=1$ to that with $p=1$ and $q=2$:
\begin{align}
&\phi_2 \frac{\fn{\varepsilon_e}{\omega}+(d-1)\varepsilon_1}{\fn{\varepsilon_e}{\omega} -\varepsilon_1} + \phi_1 \frac{\fn{\varepsilon_e}{\omega}+(d-1)\varepsilon_2}{\fn{\varepsilon_e}{\omega} -\varepsilon_2}
\label{eq:series_self-consistent} \\
=& 
2-d - \sum_{n=2}^\infty \qty[\frac{\fn{A_n^{(2)}}{k_1}}{\phi_2}{\beta_{21}}^{n-2} + \frac{\fn{A_n^{(1)}}{k_2}}{\phi_1}{\beta_{12}}^{n-2}],  \nonumber
\end{align}
where the term $(2-d)$ on the right-hand side is obtained from ${\beta_{21}}^{-1} + {\beta_{12}}^{-1}$.
Since the exact expansion \eqref{isotropic-strong} is independent of choice of reference phase and because the phase-invesrion symmetry places each phase on the same footing, we can write the effective dielectric constant independent of the reference phase, i.e., $\fn{\varepsilon_e}{\omega} = \fn{\varepsilon_e}{k_1(\omega),\omega} = \fn{\varepsilon_e}{k_2 (\omega),\omega}$.
Truncating Eq. \eqref{eq:series_self-consistent} at the two-point level and solving it in $\fn{\varepsilon_e}{\omega}$ yields the following approximation:
\begin{align}
\fn{\varepsilon_e}{\omega} =& \frac{\varepsilon_1 +\varepsilon_2}{2} +\frac{1}{2\fn{\mathcal{A}_2}{\omega}}\Big(-d\qty(\varepsilon_1\phi_2 + \varepsilon_2 \phi_1) 
\label{eq:strong-contrast-approx+SC} \\
&+ \big\{4 \fn{\mathcal{A}_2}{\omega} [d-\fn{\mathcal{A}_2}{\omega}]\varepsilon_1\varepsilon_2 \nonumber \\
& +\qty[(\varepsilon_1+\varepsilon_2)\fn{\mathcal{A}_2}{\omega} -d(\phi_2\varepsilon_1 + \phi_1\varepsilon_2)  ]^2 \big\}^{1/2}\Big), \nonumber 
\end{align}
where $\fn{\mathcal{A}_2}{\omega} \equiv d-1+\fn{A_2^{(1)}}{k_2}/\phi_1 + \fn{A_2^{(2)}}{k_1}/\phi_2 $, and we choose the physically meaningful solution from the quadratic equation.
Note that in the static limit, $\fn{\mathcal{A}_2}{\omega}$ converges to $d-1$, and thus Eq. \eqref{eq:strong-contrast-approx+SC} reduces to the well-known self-consistent formula \cite{To02a}.

Examples of phase-inversion symmetric media include the $d$-dimensional {\it random checkerboard} \cite{To02a}
and what has been called {\it  Debye random media} \cite{De57,Ye98a,Ma20}. Here we apply the approximation (\ref{eq:strong-contrast-approx+SC}) to Debye random media, which
are defined by their autocovariance function \cite{Ye98a,Ma20}
\begin{equation}\label{eq:S2-Debye}
\chi_{_V}(r) = \phi_1 \phi_2 e^{-r / r_0},
\end{equation}
where a positive quantity $r_0$ represents a characteristic length scale.
Here, we take $r_0 = a/2$.

Figure \ref{fig:phase-inversion} compares the effective dynamic
dielectric constant for  3D Debye random media to that of 3D equilibrium hard spheres
as predicted from the strong-contrast approximations devised, respectively, for phase-inversion symmetric media [Eq. (\ref{eq:strong-contrast-approx+SC})] and dispersions of particles [Eqs. \eqref{eq:modified-strong-contrast-approximation_2pt}], which do not have such a symmetry.
We see that Debye random media are
more lossy than equilibrium packings.

\begin{figure}[h]
\includegraphics[width=0.45\textwidth]{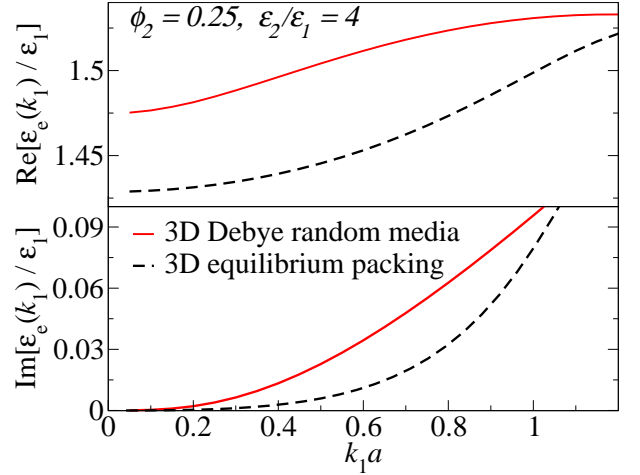}
\caption{Evaluation of the effective dielectric constant $\varepsilon_e$ as a function of the dimensionless wave number $k_1a$ for 3D equilibrium packings and 3D Debye random media of volume fraction $\phi_2=0.25$ and contrast ratio $\varepsilon_2 /\varepsilon_1 = 4$.
Here $k_1 \equiv \sqrt{\varepsilon_1}\omega/c$ is the wave number in the matrix (reference) phase 1, and $a$ is particle radius.
\label{fig:phase-inversion} }
\end{figure}

\begin{acknowledgments}
We thank Z. Ma, M. Klatt, Y. Chen, and L. Dal Negro for very helpful discussions.
The authors gratefully acknowledge  the support of Air Force Office of
Scientific Research Program on Mechanics of Multifunctional Materials and
Microsystems under Grant No. FA9550-18-1-0514.
\end{acknowledgments}

%

\end{document}